\documentclass[10pt, twocolumn, twoside]{IEEEtran}
\usepackage{graphicx}
\usepackage{amsmath}
\usepackage{caption}
\usepackage{subcaption}
\usepackage{siunitx}
\usepackage{amssymb}
\usepackage{placeins}
\usepackage{lettrine}
\usepackage{multicol, blindtext}
\usepackage{bbm}
\usepackage[multiple]{footmisc}
\usepackage[linesnumbered,ruled,vlined]{algorithm2e}
\usepackage[noend]{algorithmic}
\usepackage[geometry]{ifsym}
\usepackage{ifsym}
\usepackage{bbm}
\usepackage{amscd}
\usepackage{float}
\usepackage{dsfont}
\usepackage{amsmath}
\usepackage{epsfig}
\usepackage{psfrag}
\usepackage{rotating}
\usepackage{amsfonts}
\usepackage{url}
\usepackage{color}
\usepackage{epstopdf}
\usepackage{amsthm}
\usepackage{pgfplots}
\usepackage{authblk}
\usepackage{tikz}
\usetikzlibrary{arrows,shapes,decorations,backgrounds}
\makeatletter

\newcommand{\Rmnum}[1]{\expandafter\@slowromancap\romannumeral #1@}
\makeatother

\newcommand\blfootnote[1]{%
  \begingroup
  \renewcommand\thefootnote{}\footnote{#1}%
  \addtocounter{footnote}{-1}%
  \endgroup
}
\newtheorem{theorem}{Theorem}
\newtheorem{corollary}{Corollary}[theorem]
\newtheorem{lemma}{Lemma}[theorem]
\newtheorem{rem}{Remark}
\newtheorem{proposition}{Proposition}
\DeclareMathOperator{\interior}{int}

\ifCLASSINFOpdf
\else
\fi

\begin{document}

%\title{Modified Throughput Optimal CSMA Scheme \\ For Energy Efficiency Improvement}
%
%
%% author names and affiliations
%% use a multiple column layout for up to three different
%% affiliations
%\author{\IEEEauthorblockN{Ali Maatouk, Mohamad Assaad and Anthony Ephremides}
%\IEEEauthorblockA{TCL Chair on 5G, Laboratoire des Signaux et Systemes (L2S, CNRS)\\CentraleSup\'elec , Gif-sur-Yvette, France\\
%Email: Ali.Maatouk@supelec.fr, Mohamad.Assaad@centralesupelec.fr}
%}

\title{Energy Efficient and Throughput\\ Optimal CSMA Scheme}
\author[*]{Ali Maatouk}
\author[*]{Mohamad Assaad}
\author[$\dagger$]{Anthony Ephremides}
\affil[*]{TCL Chair on 5G, Laboratoire des Signaux et Syst\`emes, CentraleSup\'elec, Gif-sur-Yvette, France }
\affil[$\dagger$]{ECE Dept., University of Maryland, College Park, MD 20742}


%Telephone: (800) 555--1212, Fax: (888) 555--1212}


\maketitle
\blfootnote{This work has been performed in the framework of the Horizon 2020 project ONE5G (ICT-760809) receiving funds from the European Union.}
\begin{abstract}
Carrier Sense Multiple Access (CSMA) is widely used as a Medium Access Control (MAC) in wireless networks due to its simplicity and distributed nature. This motivated researchers to find CSMA schemes that achieve throughput optimality. In 2008, it has been shown that a simple CSMA-type algorithm is able to achieve optimality in terms of throughput and has been given the name ``adaptive" CSMA. Lately, new technologies emerged where prolonged battery life is crucial such as environment and industrial monitoring. This inspired the foundation of new CSMA based MAC schemes where links are allowed to transition into sleep mode to reduce the power consumption. However, throughput optimality of these schemes was not established. This paper therefore aims to find a new CSMA scheme that combines both throughput optimality and energy efficiency by adapting to the throughput and power consumption needs of each link. This is done by controlling operational parameters such as back-off and sleeping timers with the aim of optimizing a certain objective function. The resulting CSMA scheme is characterized by being asynchronous, completely distributed and being able to adapt to different power consumption profiles required by each link while still ensuring throughput optimality.
The performance gain in terms of energy efficiency compared to the conventional adaptive CSMA scheme is demonstrated through computer simulations.
\end{abstract}


% \fi
\IEEEpeerreviewmaketitle

\section{Introduction}
\lettrine{T}{hroughput} optimality for network scheduling, which is, by definition, the ability to withstand any arrival rates inside the capacity region of the network \cite{5340575}, has been thoroughly investigated in the literature. Tassiulas and Ephremides introduced in their seminal paper \cite{182479} the maximal-weight scheduling (MWS), a throughput optimal scheduling scheme. Despite
its provable optimality, this centralized scheme requires solving the maximum-weight independent set, which is an NP-Hard problem in general. With the aim of overcoming this complexity, a surge of papers have been published. The efforts were divided into two classes: the first one aimed to propose low complexity algorithms for certain interference models (e.g. \cite{4215647} and the references therein) while the other focused on proposing low complexity algorithms that guarantee a portion of the capacity region \cite{4161913} or a worst-case performance \cite{6406138}.
These approaches suffer from message passing, which can create a lot of congestion especially in the case of high links density. 
%On the other hand, another class of simple and distributed algorithms is the Carrier  Sense Multiple Access (CSMA), which is one of the most popular random access protocols. CSMA is referred to as the class of algorithms in which a link senses the medium and transmit a packet only if the medium is sensed idle. The reason for its widespread is the fact that it is simple and completely distributed \cite{6406138}. Classical CSMA such as the one adopted in IEEE 802.11 suffers however from throughput sub-optimality. This issue has been solved in a series of work where it was shown that throughput optimality can be achieved using fully distributed algorithms (e.g. \cite{5340575}\cite{shah2012}\cite{6097082}). More particularly, it was shown in \cite{5340575} that adaptive CSMA maximizes the throughput of the system for a general interference model. 

Since the work on Max-Weight, it
has been a long-standing open problem in the research community to find simple random
access schemes that achieve full optimality for a general interference model without any message passing. It wasn't until 2008 that a simple \emph{capacity achieving} Carrier Sense Multiple Access 
(CSMA) scheme, called the ``adaptive CSMA", was introduced \cite{5340575}. CSMA is a class of simple and distributed  algorithms and is considered one of the most popular 
random access protocols. In this class of algorithms, a node senses the medium and transmits a packet only if the medium is sensed idle. A suboptimal CSMA scheme 
has been already adopted for practical applications (e.g. CSMA is the basic medium access
algorithm in IEEE 802.11). Since the work of \cite{5340575}, more and more
research interests in this so-called optimal CSMA area have been taken in the community \cite{6406138}. For instance, the authors in \cite{6097082} 
proposed a distributed queue length based CSMA protocol that was proven to be throughput optimal. Interestingly, none of the previous work on optimal CSMA have investigated the power consumption aspect of these schemes. In fact, the adaptive CSMA suffers from high energy consumption: when in back-off stages, continuous sensing of the environment by each link is required 
to spot any interfering transmissions from its neighbors which would result in an inevitable power consumption \cite{5340575}. With long battery life being a strict 
requirement for emerging technologies, this line of work is of broad interest.

On the other hand, low power consumption has been the core focus in the design of Medium Access Control (MAC) protocols in Wireless Sensor Networks (WSN) \cite{1632658}. These protocols 
mainly rely on the idea of \emph{duty cycling}. Duty cycling is an effective method of reducing energy dissipation in Wireless Sensor Networks (WSNs) where a link is 
periodically placed into sleep mode. A large amount of MAC protocol solutions that are based on duty cycling were therefore proposed for WSN in the literature \cite
{1632658}. For instance, Sensor-MAC, one of the most famous low-energy MAC protocols, was proposed in \cite{1019408}. It is 
based on sleep schedules that are managed by virtual clusters. Other MAC protocols were also introduced to further enhance the performance \cite
{vanDam:2003:AEM:958491.958512}. These protocols however suffer from message passing in order to maintain synchronization. WiseMac is another MAC 
protocol that was introduced to the WSN literature in \cite{1358412}. In this protocol, all nodes wake up regularly using a common sleeping period but with different sleep schedule offsets. The clock drift can be compensated through several proposed methods (see, e.g., \cite{Sundararaman2005281}). Although there 
have been a decent number of propositions, the throughput optimality of these proposed methods was not established. 

The primary contribution of our paper is the introduction of a distributed CSMA scheme that does not include any message passing and combines both aformentioned aspects: throughput optimality and energy efficiency. The proposed scheme involves giving each link the freedom to transition between \emph{AWAKE} and \emph{SLEEP} states. However, unlike the conventional duty cycling that has been previously proposed in the literature, the sleeping duration of each link is not fixed and is calibrated with the aim of optimizing a particular objective function. In fact, we argue that by jointly controlling both the sleeping and back-off duration of each link with the aim of optimizing a certain objective function, the power consumption of each link can be reduced while \emph{still} maintaining the ability to withstand any feasible arrival rate. 
Our second and main contribution revolves around 
the use of different mathematical tools to prove vital theorems related to our scheme. In fact, the additional freedom given to each node will drastically change the CSMA model and the subsequent analysis as compared to the work in \cite{5340575} as one will see in the sequel (e.g. the proof of theorem 2 is based on totally different tools). Consequently, after appropriate analysis, 
the result will be a fully distributed MAC algorithm in which time is not slotted (hence no synchronization required) and does not suffer from any message passing or clock 
drift issues. A new parameter will reveal itself which is assigned to each link based on a satisfactory power-delay tradeoff. Moreover, as a final contribution, we provide implementations insights on our scheme. We show that the proposed scheme, due to the dynamic nature of the activity of links, achieves similar collisions performance as the adaptive CSMA while drastically reducing the power consumption which makes it appealing to be implemented in practice. The performance gain with respect to the optimal adaptive CSMA scheme is corroborated by computer simulations.

%this paper seeks to find a scheme that includes a \emph{smart} duty cycling combined with the usual adaptive CSMA in the aim to preserve throughput optimality while lowering energy consumption.
%\textit{The Markov Chain of our proposed scheme is different from that of Adaptive CSMA, which requires new theoretical performance analysis of the proposed scheme (e.g. Theorem 2).  }
The paper is organized as follows: Section \Rmnum{2} describes the system model and presents the proposed CSMA scheme. A Markov chain based analysis is provided in Section \Rmnum{3} to prove the throughput optimality of our proposed scheme and introduce the notion of energy efficiency. Insights on the implementation of our scheme in practical scenarios are given in Section \Rmnum{4}. Section \Rmnum{5} provides numerical results that corroborate the theoretical results while Section \Rmnum{6} concludes the paper.
\section{System Model and Proposed Algorithm}
\subsection{Proposed Scheme}
We consider a wireless network composed of $K$ links using the same bandwidth where each link is an (ordered) transmitter-receiver pair. In the sequel, we primarily focus on the transmitter side of each link in which we are interested in its ability to send DATA while carefully addressing its power consumption. In this framework, the receiver side of each pair can be a certain access point that is always ON (an application example being a Home Automation Network (HAN) \cite{TOSCHI201742}) or a device that is equipped with a radio-triggered wake-up mechanism \cite{1317246}. Based on that, the words ``link" and ``transmitter" will be used interchangeably in the remainder of the paper.
The proposed scheme in this paper belongs to the family of CSMA MAC protocols where transmitters in the network listen to the medium before proceeding to the transmission. More specifically, an \emph{active} transmitter (i.e. not asleep) waits for a certain duration before transmitting, called the \emph{back-off} time. While waiting, it keeps sensing the environment to spot any conflicting transmissions. If an interfering transmission is spotted, the transmitter stops immediately its back-off timer and waits for the medium to be free to resume it. Motivated by reducing the power consumption, we provide each transmitter the freedom to transition between two states:
\begin{itemize}
\item \emph{SLEEP} state: in this state, power consumption is minor and no sensing of the environment takes place (the transmitter's radio is OFF)
\item \emph{AWAKE} state: the transmitter's radio is ON and the adaptive CSMA scheme takes place as will be detailed in the sequel
\end{itemize}
The decision to either wake-up/sleep is dictated by an appropriate timer. When a transmitter decides to sleep, it picks an exponentially distributed wake-up time with mean $1/W_k$ after which it wakes up. Once the transmitter is awake, it picks an exponentially chosen sleep timer with mean $1/S_k$ after which it goes back to sleep. The motivation behind the exponential distribution assumption is the memory-less property that allows us to pursue a Markov chain based analysis as will be further explored in a future section. When the transmitter is awake, the adaptive CSMA scheme takes place:
\begin{itemize}
\item An exponentially distributed back-off timer with mean $1/R_k$ is picked
\item Continuous sensing of the channel takes place and whenever the channel is sensed idle, the back-off timer runs otherwise it is frozen. In both cases, the sleep timer keeps running
\item Once the back-off timer runs out, the sleep timer is frozen and the packet's transmission starts. 
\item Packet's transmission time is assumed to be exponentially distributed with an average channel holding time $1/H_k$
\item After successful transmission, the sleep-timer is resumed and a new back-off timer is picked for the next transmission
\end{itemize}
\subsection{System Model}
%We consider a wireless network composed of multiple links (i.e. multiple transmitters and receivers pairs) using the same bandwidth. The interference between the links is modeled by a conflict graph, which is a common model used in the literature and more precisely in the area of random access and CSMA based modeling. Assuming that we have $K$ links in the network, we associate a corresponding conflict graph G=$\{V,E\}$,  where $V$ is the set of vertices of the graph which denotes the links of the network and $E$ being the set of edges. Two vertices have an edge between them iff these two links cannot transmit at the same time. Clearly, not all links necessarily have the ability to transmit at the same time hence we define the notion of independent set of this conflict graph. We denote the independent set of this graph by $x^i\in\{0,1\}^K$ where $x_k^i=1$ if link $k$ is active in this set.
%
%With the aim of reducing power consumption, we give the ability to each node to choose either to be awake or asleep. In this context, we have $2^K$ possible awake configurations for the links. For each configuration $j$, there exist a unique conflict graph $G_j$=$\{V_j,E_j\}$. We define for each of these graphs, the set of independent sets
%$I_j$=\{$x^i$: the network is in awake configuration $j$\}.
In this section, we provide the necessary system model details to further proceed with the mathematical analysis. As a first step, we tackle the interference model. More precisely, the interference between the links is modeled by a conflict graph, which is a common model used in the literature and more precisely in the area of random access and CSMA based modeling. We recall that in the aim of reducing power consumption, we give the ability to each link to be either awake or asleep. In this context, we define the $j^{th}$ \emph{configuration state} $a^j$ as a $K$-tuplet of binary variables $a^j_k$ that indicates if link $k$ is awake (binary value 1) or asleep (binary value 0). In fact, we have $2^K$ possible configuration states for the links. For each \emph{configuration state} $a^j$, there exist a unique conflict graph $G_j$=$\{V_j,E_j\}$ where $V_j$ is the set of \emph{awake} links and $E_j$ being the set of edges. Two vertices (\emph{awake} links) have an edge between them if these two links cannot transmit at the same time. Clearly, not all links necessarily have the ability to transmit at the same time and we therefore define the notion of independent set of this conflict graph. Let $|I_j|$ be the total number of independent sets in the graph $G_j$, we denote the $i^{th}$ independent set as $x^i$. The independent set $x^i$ (also referred to as \emph{transmission state}) is a $K$-tuplet of variables $x^i_k$ that indicates if link $k$ is transmitting or not. We assume that a link transmits just one packet when it acquires the channel, i.e., $x_k^i$ is a binary variable. 
%Note that if $x_k^i=1$, this necessarily means that the associated configuration state with this transmission state must have $a_k^j=1$ since a link cannot transmit unless it is awake. 

In the following, we adopt the standard idealized CSMA assumptions firstly introduced by \cite{1096769} and adopted by the authors in \cite{5340575}:
\begin{itemize}
 \item The problem of \emph{hidden nodes} does not exist
\item Sensing is considered instantaneous, there is no sensing delay
\end{itemize}
The first condition is plausible in realistic scenarios if the range of carrier-sensing is large enough \cite{4359012}. As for the second condition, it is violated in practical systems due to the finite speed of light and the time needed for a receiver to detect the radio signals. This condition however simplifies the mathematical model, making it tractable, which can act as a starting point before considering more complicated scenarios where this condition is violated (see Section \Rmnum{4}-C for the case where this condition is violated). In addition to that, we assume that the links are always back-logged. With these idealized CSMA settings and the continuous nature of the back-off timer, collisions become mathematically impossible. This leads to tractability as a first step of the study and enables us to capture the essence of the scheduling problem without being concerned about the contention resolution problem (we tackle the contention resolution problem in Section \Rmnum{4}-C). In the next section, we present a Markov chain based analysis of our scheme to prove its throughput optimality along with a discussion on its energy efficiency.
\begin{figure*}[ht]
\centering
\includegraphics[height=4.7cm]{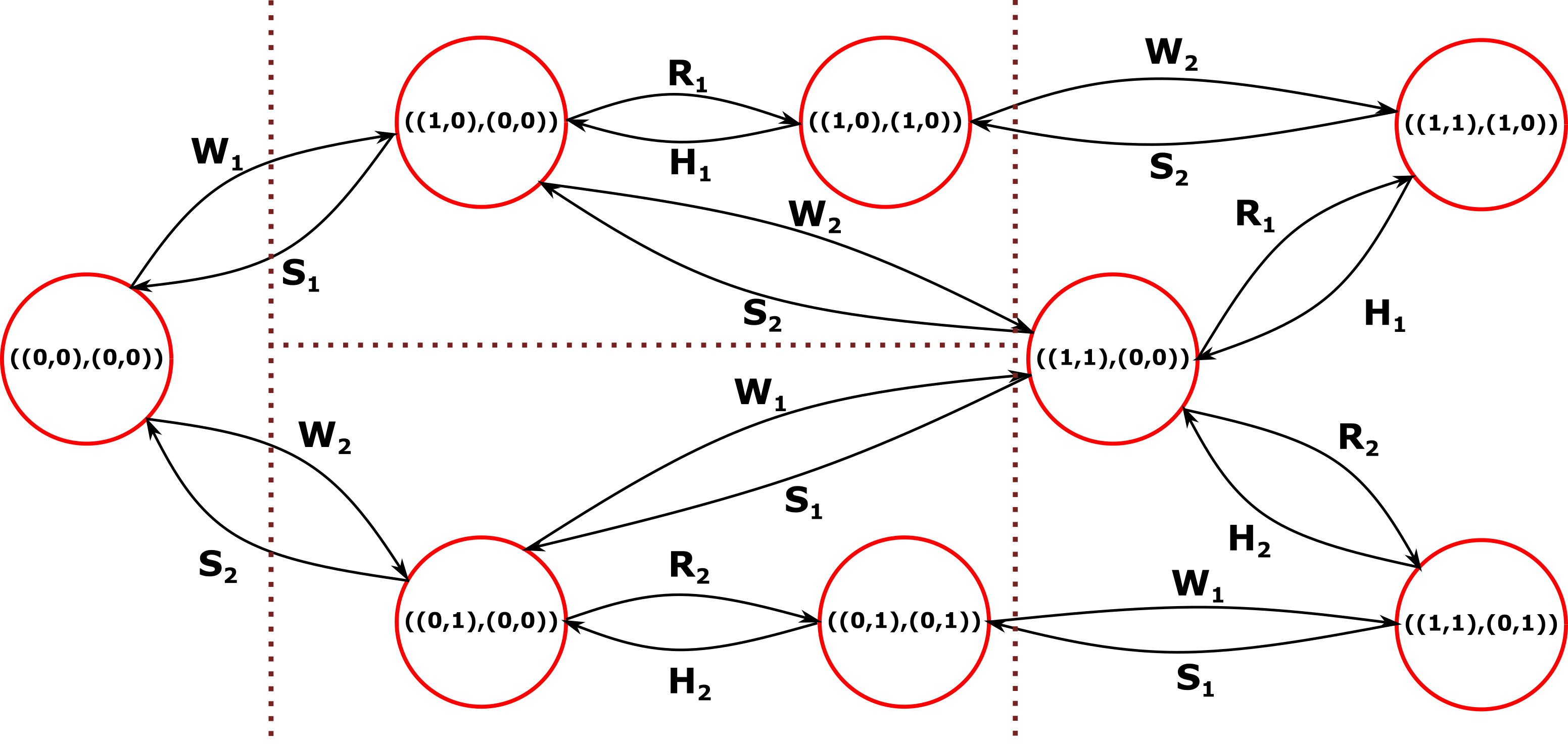}
\caption{Two interfering links Markov chain}
\label{fig:MC example}
\end{figure*}

%The words nodes, links and users are used interchangeably throughout the paper. 

%Taking these assumptions into consideration, we can now state the way our mechanism works. The link can be in one of the following states:

%Combining the previous assumptions with time continuity of the back-off timer, the probability of collision is null.

%\begin{rem}
%One can see that our framework provides more degrees of freedom as compared to the conventional CSMA since new parameters, for instance $1/W_k$ and $1/S_k$, are introduced.  Note that increasing the back-off timer in the conventional is not sufficient to provide energy efficiency as the link must perform a sensing of the medium. Furthermore, in the adaptive CSMA [4],  it is worth mentioning that our Markov Chain is more complex as compared to the existing CSMA models, and therefore the analysis requires new techniques (e.g. Theorem 2 here cannot be proved using standard convex optimization arguments as in [4]).   
%\end{rem}
\color{black}
\section{Markov chain Based Analysis}
\subsection{Proposed Markov Chain}
Let us consider the 2D continuous time stochastic process $\{\big(\boldsymbol{A}(t),\boldsymbol{X}(t)\big): t\geq 0\}$
where $\boldsymbol{A}(t)\in\{0,1\}^K$ and $\boldsymbol{X}(t)\in\{0,1\}^K$ denote the configuration and transmission states of the network at time instant $t$ respectively.
%Due to the introduction of a new degree of freedom for each node, the Markov chain proposed in \cite{5340575} would not fit our analysis. 
%As a matter of fact, our proposed Markov Chain is essentially an extension of the one proposed in \cite{5340575}. 
By adopting our proposed scheme, $\{\big(\boldsymbol{A}(t),\boldsymbol{X}(t)\big): t\geq 0\}$ becomes a Markovian process and the network can be therefore represented by a 2D continuous time Markov chain where each state in the chain is made of two components $(a^j,x^i)$ that were previously explained.
%\textit{Configuration State $a^j$:} A $K$-tuplet of binary variables $a_k$ that indicates if the node k is awake (binary value 1) or asleep (binary value 0).\\
%\textit{Transmission State $x^i$:} A $K$-tuplet of variables $x_k$ that indicates if the node $k$ is transmitting or not. We assume that a node transmit just one packet when it acquires the channel hence $x_k^i$ is a binary variable. To note that if $x_k^i=1$, this necessarily means that the associated configuration state with this transmission state must have $a_k^j=1$ since a node cannot transmit unless it is awake.

In the sequel, we define the \textit{transmission aggressiveness} and the \textit{waking-up aggressiveness} as $\boldsymbol{r}\in \mathbb{R}^K$ and $\boldsymbol{\rho}\in \mathbb{R}^K$ where 
$r_k=\log(R_k/H_k)$ and $\rho_k=\log(W_k/S_k)$ respectively.

\begin{proposition}The 2D continuous time Markov chain is irreducible, time-reversible and is fully characterized by the \textit{transmission aggressiveness} $\boldsymbol{r}$ and \textit{waking-up aggressiveness} $\boldsymbol{\rho}$. Moreover, this chain admits $\pi(a^j,x^i;\boldsymbol{r},\boldsymbol{\rho})$ as stationary distribution for any feasible state $(a^j,x^i)$ (i.e. $\lim_{t\to\infty}P\big(\big(\boldsymbol{A}(t)=a^j,\boldsymbol{X}(t)=x^i\big)\big)=\pi(a^j,x^i;\boldsymbol{r},\boldsymbol{\rho})$ ) where:
\begin{equation}
\label{Stationary}
\pi(a^j,x^i;\boldsymbol{r},\boldsymbol{\rho})=\frac{\exp\Big({\sum\limits_{k=1}^{K}a_k^j\rho_k}\Big)\exp\Big({\sum\limits_{k=1}^{K}x_k^ir_k}\Big)}{C(\boldsymbol{r},\boldsymbol{\rho})}
\end{equation}
and $C(\boldsymbol{r},\boldsymbol{\rho})$ is a normalization factor:
\begin{equation}
C(\boldsymbol{r},\boldsymbol{\rho})=\sum_{j=1}^{2^K}\exp\Big({\sum_{k=1}^{K}a_k^j\rho_k}\Big)\sum_{i=1}^{|I_j|}\exp\Big({\sum_{k=1}^{K}x_k^ir_k}\Big)
\label{normalizationfactor}
\end{equation}
\label{stationarydist}
\end{proposition}
\begin{IEEEproof} It is sufficient to show that the preceding distribution verifies the detailed balance equations \cite{norris_1997}.\\
\textit{\textbf{Step 1:}}
Consider the following two states $(a^j,x^i)$ and $(a^j+\boldsymbol{e_k},x^i)$ where $a_k^j=0$ and $\boldsymbol{e_k}$ represents the canonical vector in $\mathbb{R}^{K}$ ($(\boldsymbol{e_k})_{l: 1\leq l\leq K}=(\delta_{kl})_{1\leq l\leq K}$ where $\delta_{kl}$ is the Kronecker delta function). It can be verified that:
\begin{equation*}
\frac{\pi(a^j+\boldsymbol{e_k},x^i;\boldsymbol{r},\boldsymbol{\rho})}{\pi(a^j,x^i;\boldsymbol{r},\boldsymbol{\rho})}=\exp(\rho_k)
\end{equation*}
\textit{\textbf{Step 2:}} Consider the two states $(a^j,x^i)$ and $(a^j,x^i+\boldsymbol{e_k})$ where $a_k^j=1$ (the link has to be awake to be able to transmit), $x_k^i=0$ and $x_{k'}^i=0$ $\forall  k' \in N_j(k) $ where the neighboring set of link $k$ at configuration state $a^j$ is defined as $N_j(k)=\{k': (k,k')\in E_j\}$. It can also be verified that:
\begin{equation*}
\frac{\pi(a^j,x^i+\boldsymbol{e_k};\boldsymbol{r},\boldsymbol{\rho})}{\pi(a^j,x^i;\boldsymbol{r},\boldsymbol{\rho})}=\exp(r_k)
\end{equation*}
\end{IEEEproof}
%\begin{rem}
%We recall that if the stochastic process $\{\big(\boldsymbol{A}(t),\boldsymbol{X}(t)\big): t\geq 0\}$ is time-reversible when the states $(a^j,x^i)$ are independent for all links then it is also time-reversible when restricted to the case where the states of conflicting transmissions and transmission while being asleep are forbidden. Moreover, the invariant distribution of the restriction is the renormalized stationary distribution of the stochastic process $\{\big(\boldsymbol{A}(t),\boldsymbol{X}(t)\big): t\geq 0\}$. The normalization factor of this invariant distribution is nothing but the factor depicted in (\ref{normalizationfactor}).
%\end{rem}
%\begin{figure*}[ht]
%\centering
%\includegraphics[height=3cm]{"Diagram".png}
%\caption{Time-line of the proposed CSMA scheme}
%\label{fig:timeline}
%\end{figure*}
%
%\begin{figure*}[ht]
%\centering
%\begin{subfigure}{0.5\textwidth}
%  \centering
%  \includegraphics[width=.8\linewidth]{myplot.eps}
%  \caption{Comparison of sum spectral efficiency vs. SNR}
%  \label{rate}
%\end{subfigure}%
%\begin{subfigure}{0.5\textwidth}
%  \centering
%  \includegraphics[width=0.8\linewidth]{fairness.eps}
%  \caption{Comparison of Jain's fairness index vs. SNR}
%  \label{fairness}
%\end{subfigure}
%\caption{Comparison of scheduling schemes}
%\end{figure*}
To further clarify the model, a simple example of two interfering links is taken in Fig. \ref{fig:MC example}. The dashed lines partition the chain in 4 regions where in each, the 2D states share the same \emph{configuration state}. In the first partition, $a^1=(0,0)$ and all links are asleep and the only feasible transmission state is $x^1=(0,0)$. As for the partition where $a^2=(1,0)$, only link $1$ is awake and therefore the only possible transmission states are $x^1=(0,0)$ and $x^2=(1,0)$. It is completely symmetrical for the case of $a^3=(0,1)$. As for the final partition where $a^4=(1,1)$, both links are awake. Since the two links are interfering, the conflict graph in this partition is made of two vertices mapped to the \emph{awake} links with an edge existing between them. The independent sets of this conflict graph are $x^1=(0,0),x^2=(1,0),$ and $x^3=(0,1)$. In other words, the transmission state $x^4=(1,1)$ is unfeasible.
\begin{rem}
One can clearly see how our proposed chain is much more general than the standard CSMA Markov chain by observing that the partition of the chain where $a^j=(1,\ldots,1)$ is nothing but the CSMA Markov chain firstly introduced in \cite{1096769} and adopted in \cite{5340575}.
\end{rem}
\noindent Using the 2D Markov chain, we can determine several key performance indicators of our scheme. First, the throughput achieved by user $k$ is simply the amount of time the chain is in a state where link $k$ is both awake and transmitting. This can be expressed in the following way:
\begin{equation}
\label{Throughput}
s_k(\boldsymbol{r},\boldsymbol{\rho})=\sum_{j=1}^{2^K}a_k^j\sum_{i=1}^{|I_j|}x_k^i\pi(a^j,x^i;\boldsymbol{r},\boldsymbol{\rho})
\end{equation}
Moreover, we can define the \emph{awake duration} of each link $k$ as the amount of time the chain is in a state where $a_k^j=1$. In other words: 
\begin{equation}
\label{awakeduration}
\hat{f_k}(\boldsymbol{r},\boldsymbol{\rho})=\sum_{j=1}^{2^K}a_k^j\sum_{i=1}^{|I_j|}\pi(a^j,x^i;\boldsymbol{r},\boldsymbol{\rho})
\end{equation}
\subsection{Optimality Analysis}

%
%We consider that that the arrival rate vector $\boldsymbol{\lambda}\in \mathbb{R}^{K}_{+}$ is normalized to the capacity, i.e. $\lambda_k\leq1 \:\: \forall k$. In this case, $\lambda_k$ can be seen as the portion of time the chain should be in a state where user $k$ is both awake and transmitting to serve the arrival traffic. Hence a rate is said to be feasible if it can be written as a joint probability distribution $\boldsymbol{p}$ over our Markov chain space. Therefore, for any feasible rate $\boldsymbol{\lambda}$, we have:
A scheduling scheme is said to be throughput optimal if it can support any \emph{feasible} incoming rate $\boldsymbol{\lambda}\in \mathbb{R}^{K}_{+}$, i.e., it stabilizes all the queues within the network whenever it is possible to do so. Knowing that packets arrive randomly with an average rate $\lambda_k$, the queue of link $k$ is defined as the number of packets buffered by the transmitter side of the link. A buffered packet leaves the system whenever link $k$ captures the channel for an exponential holding time of average $1/H_k$. As it was shown, our scheme allows each link $k$ to achieve a throughput of $s_k(\boldsymbol{r},\boldsymbol{\rho})$. We will be able to prove that our proposed scheme is indeed throughput optimal. However, before proceeding to our proof of throughput optimality, we first define the set of feasible rates $\Lambda$. Without loss of generality, we consider that the arrival rate vector $\boldsymbol{\lambda}\in \mathbb{R}^{K}_{+}$ is normalized to the capacity of each link, i.e. $\lambda_k\leq1 \:\: \forall k$. In this case, $\lambda_k$ can be seen as the portion of time the chain should be in a state where user $k$ is both awake and transmitting to serve the arrival traffic. Hence, a rate is said to be feasible if it can be written as a joint probability distribution $\boldsymbol{p}$ over our Markov chain space. Therefore, for any feasible rate $\boldsymbol{\lambda}$, we have:
\begin{equation}
 \setlength{\belowdisplayskip}{10pt} \setlength{\belowdisplayshortskip}{10pt}
\setlength{\abovedisplayskip}{10pt} \setlength{\abovedisplayshortskip}{10pt} 
\label{Throughput}
\lambda_k=\sum_{j=1}^{2^K}\sum_{i=1}^{|I_j|}p_{ij}a_k^jx_k^i
\end{equation}
and $\boldsymbol{p}$ satisfies:
\begin{equation*}
p_{ij}\geqslant0 \:\: \forall (i,j) \quad \textrm{and} \quad \sum_{j=1}^{2^K}\sum_{i=1}^{|I_j|}p_{ij}=1
\end{equation*}
We can therefore define $\Lambda$ as follows:
\begin{equation*}
\Lambda=\Big\{\boldsymbol{\lambda}\in\mathbb{R}^{K}_{+}|\:\: \exists \boldsymbol{p} : \lambda_k=\sum_{j=1}^{2^K}\sum_{i=1}^{|I_j|}p_{ij}a_k^jx_k^i\Big\}
\end{equation*}
where $\boldsymbol{p}$ verifies: 
\begin{equation*}
\begin{cases}
p_{ij}\geqslant0 \:\: \forall (i,j) \\
\sum\limits_{j=1}^{2^K}\sum\limits_{i=1}^{|I_j|}p_{ij}=1
\end{cases}
\end{equation*}
Furthermore, we introduce a new parameter $\boldsymbol{f}\in \mathbb{R}^{K}_{+}$, which we will refer to as the \emph{awake} vector, as follows:
%\begin{equation}
%f_k=\sum_{j=1}^{2^K}a_k^j\sum_{i=1}^{|I_j|}p_{ij} \quad \forall 1\leq k\leq K
%\end{equation}
\begin{equation}
f_k=\sum_{j=1}^{2^K}a_k^j\alpha_j=\mathbb{E}(a_k)\quad \textrm{and} \quad \alpha_j=\sum_{i=1}^{|I_j|}p_{ij}
\end{equation}
$\alpha_j$ represents the portion of time required for the network to be at configuration state $a^j$ while $f_k$ can be seen as the required awake duration of each link $k$, both as dictated by the arrival rate vector's joint probability distribution $\boldsymbol{p}$. This newly introduced parameter will be key to the energy consumption aspect of our scheme as will be further detailed in the sequel. After these proper definitions, the throughput optimality of our proposed scheme is depicted in the following theorem.
\begin{theorem}
For any arrival rate $\boldsymbol{\lambda} \in \Lambda$,  there exist $(\boldsymbol{r}^*,\boldsymbol{\rho}^*)$ such that $s_k(\boldsymbol{r}^*,\boldsymbol{\rho}^*)=\lambda_k \:\: \forall k $. Moreover, $\hat{f_k}(\boldsymbol{r}^*,\boldsymbol{\rho}^*)= f_k \:\: \forall k$.
\label{existence}
\end{theorem}
Before proceeding to the proof, we first present our approach. For any feasible arrival rate $\boldsymbol{\lambda} \in \Lambda$, our goal can be summarized as calibrating the parameters $(\boldsymbol{r},\boldsymbol{\rho})$ in a way to make our CSMA Markov chain's stationary distribution as close as possible to $\boldsymbol{p}$. This is equivalent to reducing the distance between these two distributions. Different measures between probability distributions exist but in our paper we adopt the \textit{Kullback-Leibler} divergence as in \cite{5340575}. This approach was firstly introduced by the authors in \cite{5340575} and was motivated by results on the theory of Markov random fields \cite{Wainwright:2008:GME:1498840.1498841}. In fact, when minimizing the \textit{Kullback-Leibler} divergence between a certain joint distribution and a product-form joint distribution, some particular marginal distributions induced by these two distributions are equal. This approach fits our framework since the joint distribution for us is nothing but $\boldsymbol{p}$ and the product-form distribution is the stationary distribution of our proposed CSMA scheme depicted in Proposition \ref{stationarydist}.
\begin{rem}
Although our throughput optimality's approach is similar to that of \cite{5340575}, the subsequent analysis is different due to the introduction of new parameters. In fact, the structure of the optimization problem leads us to introduce new variables in order to simplify the analysis through convex optimization tools. Moreover, the results on the finiteness of the optimal parameters in Theorem \ref{finite}, which are pivotal to the work, will be based on different mathematical tools as it will be explained in the sequel.
\end{rem}
\color{black} 
\begin{IEEEproof}
\noindent To pursue our optimality analysis, we formulate the optimization problem as follows:
\begin{equation*}
\setlength{\belowdisplayskip}{0pt} \setlength{\belowdisplayshortskip}{0pt}
\setlength{\abovedisplayskip}{0pt} \setlength{\abovedisplayshortskip}{0pt} 
\begin{aligned}
& \underset{\boldsymbol{r},\boldsymbol{\rho}}{\text{minimize}}
& & D(\boldsymbol{p}\|\pi(\boldsymbol{r},\boldsymbol{\rho}))=\sum_{j=1}^{2^K}\sum_{i=1}^{|I_j|}p_{ij}\log\Big(\frac{p_{ij}}{\pi(a^j,x^i;\boldsymbol{r},\boldsymbol{\rho})}\Big)
\end{aligned}
\end{equation*}
The objective function can be reduced through several steps:
\begin{multline*}
\sum_{j=1}^{2^K}\sum_{i=1}^{|I_j|}p_{ij}\log\Big(\frac{p_{ij}}{\pi(a^j,x^i;\boldsymbol{r},\boldsymbol{\rho})}\Big)=\sum_{j=1}^{2^K}\sum_{i=1}^{|I_j|}p_{ij}\log(p_{ij})\\-\sum_{j=1}^{2^K}\sum_{i=1}^{|I_j|}p_{ij}\log(\pi(a^j,x^i;\boldsymbol{r},\boldsymbol{\rho}))=\sum_{j=1}^{2^K}\sum_{i=1}^{|I_j|}p_{ij}\log(p_{ij})\\-\sum_{j=1}^{2^K}\sum_{i=1}^{|I_j|}p_{ij}\Big(\sum_{k=1}^{K}x_k^ir_k+\sum_{k=1}^{K}a_k^j\rho_k-\log(C(\boldsymbol{r},\boldsymbol{\rho}))\Big)
\end{multline*}
The first term being independent of $(\boldsymbol{r},\boldsymbol{\rho})$, the problem can be reformulated as minimizing the following function:
\begin{multline*}
F(\boldsymbol{r},\boldsymbol{\rho})=-\sum_{k=1}^{K}\sum_{j=1}^{2^K}\sum_{i=1}^{|I_j|}p_{ij}x_k^ir_k-\sum_{k=1}^{K}\sum_{j=1}^{2^K}\sum_{i=1}^{|I_j|}p_{ij}a_k^j\rho_k \\
+\log\Big(\sum_{j=1}^{2^K}\exp\big({\sum_{k=1}^{K}a_k^j\rho_k}\big)\sum_{i=1}^{|I_j|}\exp\big({\sum_{k=1}^{K}x_k^ir_k}\big)\Big)
\end{multline*}
The first term can be further reduced to an easier form since $x_k^i$ cannot be equal to 1 unless link $k$ is actually awake. Therefore, multiplying by the binary variable $a_k^j$ does not change the value of this term:
\begin{equation*}
\begin{aligned}
\sum_{k=1}^{K}\sum_{j=1}^{2^K}\sum_{i=1}^{|I_j|}p_{ij}x_k^ir_k=\sum_{k=1}^{K}\sum_{j=1}^{2^K}\sum_{i=1}^{|I_j|}p_{ij}a_k^jx_k^ir_k=\sum_{k=1}^{K}\lambda_kr_k
\end{aligned}
\end{equation*}
As for the second term, we have:
\begin{equation*}
\begin{aligned}
\sum_{k=1}^{K}\sum_{j=1}^{2^K}\sum_{i=1}^{|I_j|}p_{ij}a_k^j\rho_k=\sum_{k=1}^{K}\sum_{j=1}^{2^K}a_k^j\rho_k\sum_{i=1}^{|I_j|}p_{ij}=\sum_{k=1}^{K}f_k\rho_k
\end{aligned}
\end{equation*}
In order to solve this problem in a simple way using convex optimization tools, we introduce the new variables $t_{ji}$ in the following manner:
\begin{equation}
\begin{aligned}
& \underset{\boldsymbol{r},\boldsymbol{\rho},\boldsymbol{t}}{\text{minimize}}
&& -\sum_{k=1}^{K}\lambda_kr_k-\sum_{k=1}^{K}f_k\rho_k+\log\Big(\sum_{j=1}^{2^K}\sum_{i=1}^{|I_j|}\exp(t_{ji})\Big) \\
& \text{subject to}
& & t_{ji}=\sum_{k=1}^{K}a_k^j\rho_k+\sum_{k=1}^{K}x_k^ir_k\\
&&& j = 1, \ldots, 2^K \;i = 1, \ldots, |I_j|.
\end{aligned}
\label{convexop}
\end{equation}
The objective function is made of convex/linear functions in $(\boldsymbol{r},\boldsymbol{\rho},\boldsymbol{t})$ since the log-sum-exponential function is a well known convex function \cite{Boyd:2004:CO:993483}. On top of that, our equality constraints are linear and our transformed OP is indeed convex. 
%Therefore, we can use the Lagrangian function and obtain the global optimal solution of our problem (\ref{convexop}) by the Lagrange multipliers conditions that are \emph{necessary} and \emph{sufficient} in our case \cite{Boyd:2004:CO:993483}.
Therefore, we can use the Lagrangian function and obtain the global optimal solution of our problem (\ref{convexop}) by the Karush--Kuhn--Tucker (KKT) conditions that are \emph{necessary} and \emph{sufficient} in our case \cite{Boyd:2004:CO:993483}. These conditions simplify to the Lagrange multipliers conditions due to the absence of any inequality constraints. The Lagrangian function is formulated as follows:
\begin{multline}
\mathcal{L(\boldsymbol{r},\boldsymbol{\rho},\boldsymbol{T},\boldsymbol{\mu}})=\sum_{j=1}^{2^K}\sum_{i=1}^{|I_j|}\mu_{ji}(-t_{ji}+\sum_{k=1}^{K}a_k^j\rho_k+\sum_{k=1}^{K}x_k^ir_k)\\
-\sum_{k=1}^{K}\lambda_kr_k-\sum_{k=1}^{K}f_k\rho_k+\log\Big(\sum_{j=1}^{2^K}\sum_{i=1}^{|I_j|}\exp(t_{ji})\Big)
\label{lagrangefunction}
\end{multline}
where $\mu_{ji}$ represents the dual variable corresponding to the $\{ji\}^{th}$ constraint. At the optimal point, the following condition holds:
\begin{equation}
\frac{\partial \mathcal{L(\boldsymbol{r^*},\boldsymbol{\rho^*},\boldsymbol{T^*},\boldsymbol{\mu^*}})}{\partial t_{ji}}=-\mu_{ji}^*+\frac{\exp(t_{ji}^*)}{C(\boldsymbol{r^*},\boldsymbol{\rho^*})}=0
\end{equation}
Moreover, since the optimal point has to satisfy the imposed constraint, we have the following:
%\begin{equation}
%%\frac{\partial \mathcal{L(\boldsymbol{r^*},\boldsymbol{\rho^*},\boldsymbol{T^*},\boldsymbol{\mu^*}})}{\partial \mu_{ji}}=-t_{ji}^*+\sum_{k=1}^{K}a_k^j\rho_k^*+\sum_{k=1}^{K}x_k^ir_k^*=0
%\end{equation}
\begin{equation}
t_{ji}^*=\sum_{k=1}^{K}a_k^j\rho_k^*+\sum_{k=1}^{K}x_k^ir_k^*
\end{equation}
Combining the above two conditions, we can conclude that:
\begin{equation*}
\mu_{ji}^*=\frac{\exp\Big({\sum\limits_{k=1}^{K}a_k^j\rho_k^*}\Big)\exp\Big({\sum\limits_{k=1}^{K}x_k^ir_k^*}\Big)}{C(\boldsymbol{r^*},\boldsymbol{\rho^*})}=\pi(a^j,x^i;\boldsymbol{r^*},\boldsymbol{\rho^*})
\end{equation*}
which is nothing but the stationary distribution of our chain. Therefore, replacing our dual variable by the stationary distribution of our chain in the subsequent analysis take into account both the first order condition with respect to $t_{ji}$ and the feasibility condition. Furthermore, the following condition also holds:
\begin{equation}
\frac{\partial \mathcal{L(\boldsymbol{r^*},\boldsymbol{\rho^*},\boldsymbol{T^*},\boldsymbol{\mu^*}})}{\partial \rho_k}=-f_k+\sum_{j=1}^{2^K}\sum_{i=1}^{|I_j|}\mu_{ji}^*a_k^j=0
\label{gradientrho}
\end{equation}
Using our previous conclusion on $\mu_{ji}^*$, (\ref{gradientrho}) will lead to:
\begin{equation}
-f_k+\sum_{j=1}^{2^K}a_k^j\sum_{i=1}^{|I_j|}\pi(a^j,x^i;\boldsymbol{r}^*,\boldsymbol{\rho}^*)=0
\label{eq1}
\end{equation}
%\begin{equation*}
%-f_k+\sum_{j=1}^{2^K}a_k^j\hat{\alpha_j}=0 \quad \textrm{and} \quad \hat{\alpha_j}=\sum_{i=1}^{|I_j|}\pi(a^j,x^i;\boldsymbol{r}^*,\boldsymbol{\rho^*})
%\end{equation*}
%$\hat{\alpha_j}$ is interpreted as the amount of time the chain is in configuration state $a^j$. 
This means that at the optimal point, $\hat{f_k}(\boldsymbol{r}^*,\boldsymbol{\rho}^*)=f_k \:\: \forall k$ or in other words, link $k$ is awake just as needed to be. Lastly:
\begin{equation}
\frac{\partial \mathcal{L(\boldsymbol{r^*},\boldsymbol{\rho^*},\boldsymbol{T^*},\boldsymbol{\mu^*}})}{\partial r_k}=-\lambda_k+\sum_{j=1}^{2^K}\sum_{i=1}^{|I_j|}\mu_{ji}^*x_k^i=0
\label{eq8}
\end{equation}
As it has been used before, multiplying by the binary variable $a_k^j$ does not change this value due to the fact that a link cannot transmit if it is not awake. Also, using the previous conclusions on $\mu_{ji}^*$, (\ref{eq8}) becomes:
\begin{equation}
-\lambda_k+\sum_{j=1}^{2^K}a_k^j\sum_{i=1}^{|I_j|}x_k^i\pi(a^j,x^i;\boldsymbol{r^*},\boldsymbol{\rho^*})=0
\label{eq2}
\end{equation}
Hence, at the optimal point, we have $s_k(\boldsymbol{r}^*,\boldsymbol{\rho}^*)=\lambda_k$.
\end{IEEEproof}
\begin{corollary}
Let $Q_k(t)$ represents the queue length of link $k$ at time $t$. For any arrival rate $\boldsymbol{\lambda} \in \Lambda$, there exist $(\boldsymbol{r}^*,\boldsymbol{\rho}^*)$ such that 
the queuing systems of all links are rate stable (i.e. $\lim_{t\to\infty} \frac{Q_k(t)}{t}\rightarrow 0 \:\: \forall k)$.
\end{corollary}
\begin{IEEEproof}
As previously proven in Theorem \ref{existence}, there exist $(\boldsymbol{r}^*,\boldsymbol{\rho}^*)$ such that $s_k(\boldsymbol{r}^*,\boldsymbol{\rho}^*)=\lambda_k \:\: \forall k $. This is a sufficient condition for rate stability of each link's queuing system \cite[p.17]{Neely:2010:SNO:1941130}. 
\end{IEEEproof}
Consequently, by minimizing (\ref{lagrangefunction}), we can achieve queuing rate stability of all links in the network for any arrival rate $\boldsymbol{\lambda} \in \Lambda$. Moreover, by minimizing (\ref{lagrangefunction}), the required awake duration of each link $k$ is also achieved. As seen from the expression of (\ref{lagrangefunction}), this optimization require the knowledge of the arrival vector $\boldsymbol{\lambda}$ and awake vector $\boldsymbol{f}$. This is cumbersome on the network and is unfeasible in practice. However, by observing equations (\ref{eq1}) and (\ref{eq2}), we can see that these conditions depend on local information of each link $k$. Therefore, the optimum $(\boldsymbol{r}^*,\boldsymbol{\rho}^*)$ can be achieved in a distributed manner using a simple gradient descent algorithm where each link $k$ updates its own operational parameters $r_k$ and $\rho_k$. Details concerning this implementation are presented in Section \Rmnum{4}-B.
\subsection{Energy Efficiency}
To understand the energy efficiency aspect of our proposed scheme, it is vital to answer the following question: what does the arrival rate $\boldsymbol{\lambda}$ 
truly dictate concerning the configuration states?

\begin{figure}[!ht]
\centering
\includegraphics[width=.55\linewidth]{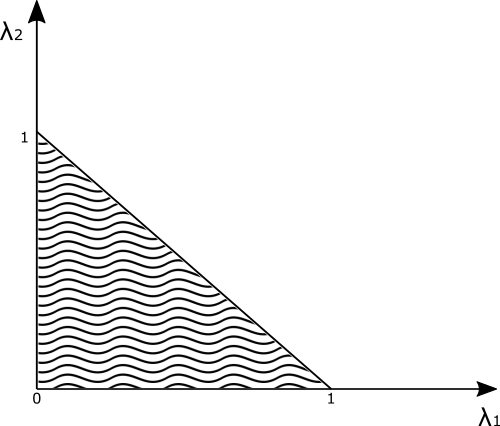}
\caption{Capacity region for the case of two interfering links}
\label{capacityreg}
\end{figure}

\begin{figure}[!ht]
\centering
\includegraphics[width=.55\linewidth]{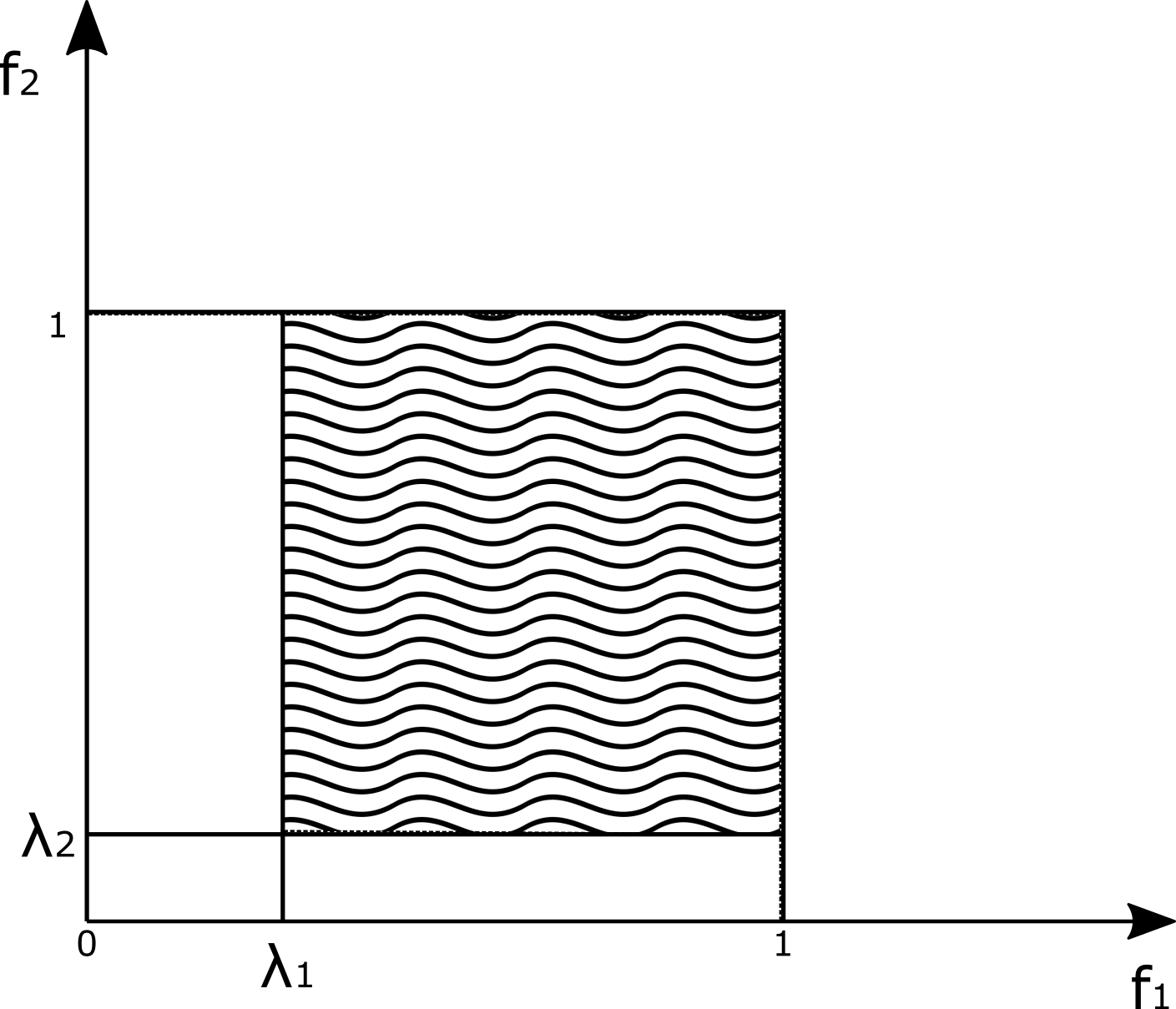}
\captionsetup{justification=centering}
\setlength{\belowcaptionskip}{-10pt}
\caption{Awake region for the case of two interfering links for a fixed $\boldsymbol{\lambda}\in \mathbb{R}^{2}_{+}$}
\label{wakingreg}
\end{figure}

To answer that, we can clearly see from the expression for $\lambda_k$, for any feasible arrival rate $\boldsymbol{\lambda}$, that the arrival rate \emph{only} dictates 
the portion of time where link $k$ is both awake and transmitting. The other states where link $k$ is either: awake and in a back-off stage or asleep, are irrelevant 
to the arrival rate and therefore the portion of time in which the chain is in those states presents for us a degree of freedom we can take advantage of. In other 
words, a feasible arrival rate $\boldsymbol{\lambda}$ has different joint probability distribution $\boldsymbol{p}$ representations each of which leading to the same $
\boldsymbol{\lambda}$ but with \emph{different} awake vector $\boldsymbol{f}$. 
To see this more clearly, for a fixed arrival rate $\boldsymbol{\lambda}$, it is straightforward that the minimal value of $f_k$ is $\lambda_k$. In this case, the joint probability distribution $\boldsymbol{p}$ verifies $p_{ij}=0$ where $a^j_k=1$ and $x^i_k=0$ (i.e. we should not be in a state where link $k$ is awake and not transmitting). Similarly, on the other extreme, the maximal value for $f_k$ is $1$. In this case, $\boldsymbol{p}$ verifies $p_{ij}=0$ where $a^j_k=0$ (i.e. states where link $k$ is asleep are forbidden). Therefore, the network is defined by two vectors ($\boldsymbol{\lambda}$,$\boldsymbol{f}$) instead of being solely defined by the arrival rate vector. Motivated by these results, we define what we will call the \emph{awake region} as the awake vector feasibility region:
%To see this more clearly, for a fixed arrival rate $\boldsymbol{\lambda}$, it is straightforward that the minimal value of $f_k$ is $\lambda_k$. The only way for a link $k$ to be awake for a fraction of time $\lambda_k$ while satisfying its throughput requirement is to instantly acquire the channel when it wakes up. On the other extreme, the maximal value for $f_k$ is $1$, i.e. link $k$ is always kept awake and it can compete for the channel to satisfy the throughput requirement of $\lambda_k$ accordingly\footnote{The special case where $f_k\longrightarrow1$ $\forall k$, we will have the configuration state $a^j=(1,\ldots,1)$ almost surely and we are back to the 
%adaptive CSMA scheme in \cite{5340575}}. Motivated by these results, we define what we will call the \emph{awake region} as the awake vector feasibility region:
\begin{equation} 
\Theta(\boldsymbol{\lambda})=\Big\{\boldsymbol{f}\in \mathbb{R}^{K}_{+}: \lambda_k\leq f_k\leq1\}
\end{equation}
We can conclude that the network is characterized by two regions rather than one: the \emph{capacity} region and the \emph{awake} region. An example is presented in Fig. \ref{capacityreg} 
and \ref{wakingreg} for the case of two interfering links.
\begin{figure*}[ht]
\centering
\includegraphics[height=2.55cm]{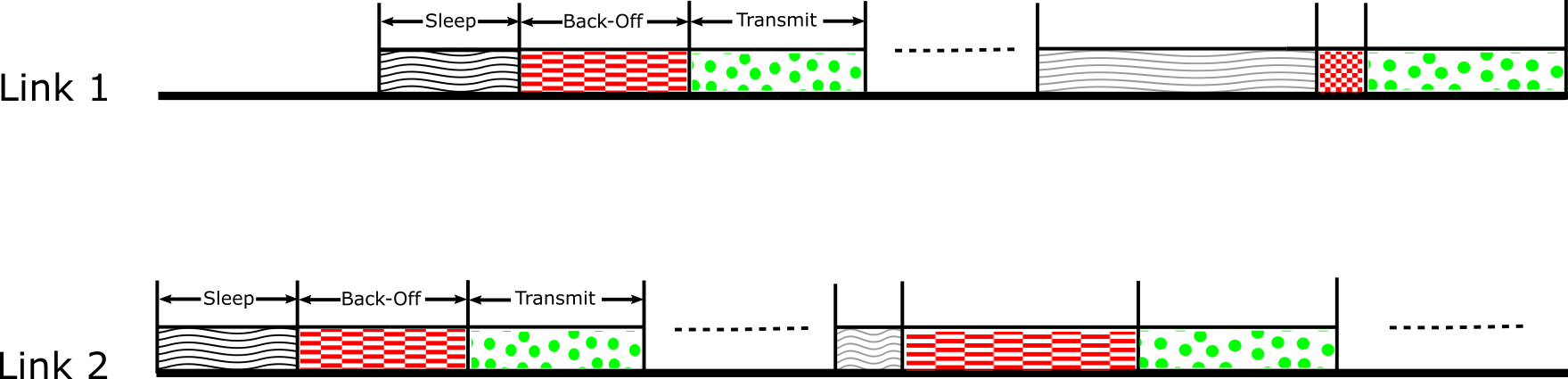}
\caption{Time-line of the proposed CSMA scheme}
\label{fig:timeline}
\vspace{-15pt}
\end{figure*}

After clarifying this, we can discuss the energy efficiency aspect of our proposed scheme. By rewriting $f_k$ in the following manner:
\begin{equation*}
f_k=\lambda_k+\omega_k
\end{equation*}
where $0<\omega_k<1-\lambda_k$, we can see that a new parameter is born. This parameter is referred to as the Power-Delay Tradeoff (PDT) parameter and is assigned to each link $k$ as will be explained in the following. In fact, when we assign a small parameter value $\omega_k$ to link $k$, the required awake time $f_k$ will be close to $\lambda_k$. By recalling the results of Theorem \ref{existence}, we know that using our scheme, link $k$ will calibrate itself to satisfy the arrival and awake time requirements. Therefore, link $k$ will be awake just as necessary ( $\hat{f_k}=f_k\thickapprox\lambda_k$) while still ensuring to satisfy its throughput requirement ($s_k=\lambda_k$). To do so, link $k$ will spend a high amount of time being asleep but when it wakes up, it is extremely aggressive on the channel (small back-off timer) and therefore power consumption is reduced. However, due to the high fraction of time spent inactive, delays are to be anticipated which would manifest in long queues lengths. On the other extreme, if the assigned parameter $\omega_k$ to link $k$ is close to its maximal value, the required awake time $f_k$ will be close to $1$. In this manner, using our scheme, link $k$ will be forced to remain almost continuously awake while being less aggressive on the channel\footnote{The special case where $f_k\longrightarrow1$ $\forall k$, all links will remain continuously awake and therefore the transition to SLEEP state does not take place for any link and we are back to the 
adaptive CSMA scheme in \cite{5340575}}. Therefore, less delays would be expected but a higher power consumption takes place. In summary, $\omega_k$ is a Power-Delay Tradeoff (PDT) parameter that is \emph{assigned} to each link $k$ depending on the desired tradeoff. The beauty of this parameter comes from the fact that IoT applications present a mixture of both delay-sensitive and delay-tolerant applications. This can be exploited by simply calibrating the parameter assignment accordingly, the following table gives a few examples:
\vspace{-10pt}
\begin{center}
\begin{tabular}{|c|c|c|}
 \hline
 \multicolumn{3}{|c|}{IoT Applications} \\
 \hline
 Category & Delay & Parameter $\omega_k$ \\
  \hline
 Emergencies  & Highly Sensitive & High\\
 E-Health  & Med. Sensitivity & Medium\\
 Environment Monitors  & Highly Tolerant &  Low\\
 \hline
\end{tabular}
 \captionof{table}{Parameter Assignment}
\end{center}
%\begin{rem}
%One can see that our framework provides more degrees of freedom as compared to the adaptive CSMA since new parameters, for instance $1/W_k$ and $1/S_k$, are introduced. In the adaptive CSMA, the overall optimal transmission aggressiveness (and therefore the optimal back-off timer mean) is just dictated by the network. This might lead to relatively large back-off timers which would result in an inevitable power loss due to continuous sensing of the environment. The motivation behind the utility of these new degree of freedom is the fact that links that sleep tend to be more aggressive on the channel once they are awake to compensate for the time spent in Sleep mode. By doing so, back-off timers are drastically shrinked which reduces power consumption. One can see in Fig. \ref{fig:timeline} an example of the case of two links where link 1 is highly constrained by its battery life unlike link 2. Link 1 therefore eventually end up in sleeping mode for large duration of time while being extremely aggressive on the network (short back-off timer) once it transitions to Awake state. On the other hand, link 2 barely sleeps and is moderately aggressive on the network. Our aim is therefore to develop a well designed sleep-wake up mechanism that would still maintain throughput optimality while dealing with this power consumption issue and this is what we are going to tackle in the following sections.
%\end{rem}

\begin{rem}
In the adaptive CSMA, the overall optimal transmission aggressiveness $\boldsymbol{r}^*$ (and therefore the optimal back-off timer) is just dictated by the network arrival rate vector $\boldsymbol{\lambda}$ and the number of links in the collision domain. This might lead to relatively large back-off timers which would result in an inevitable power loss due to the continuous sensing of the environment. For our framework, we can work around this thanks to the newly introduced parameter. In fact, by assigning a small PDT parameter, links sleep to save power but are more aggressive on the channel once they are awake to compensate for the time spent in SLEEP state and achieve the required throughput. By doing so, back-off timers are drastically reduced (i.e. less power consumption from listening to the medium) and links therefore can still achieve throughput optimality even with their ability to transition to SLEEP state. For instance, one can see in Fig. \ref{fig:timeline} an example of the case of two links where link 1 is assigned a low PDT parameter unlike link 2 that is assigned a larger one. At first, both links start with the same initial waking-up and transmission aggressiveness $(\boldsymbol{r}_0,\boldsymbol{\rho}_0)$. As the parameters converge to their optimal values, link 1 ends up in SLEEP state for a large duration of time while being extremely aggressive on the network once it transitions to AWAKE state. On the other hand, link 2 is moderately aggressive on the network.
\end{rem}
\begin{rem}
In general max-weight scheduling, the complexity-delay-stability tradeoff has been studied for wireless networks in the literature \cite{Yi:2008:CWS:1374618.1374624}. In our line of work, we are interested in CSMA based scheduling with the particularity to have energy consumption reduction while keeping the throughput optimality. The new parameter that we have just introduced $\boldsymbol{w}$, can be seen as a Power-Delay tradeoff parameter such that whatever the assigned parameter $\boldsymbol{0}<\boldsymbol{w}<\boldsymbol{1}-\boldsymbol{\lambda}$ is, the full stability region is achieved by the scheme. 
\end{rem}
\color{black}
This scheme will only be of interest if the optimum is always attained for a finite $(\boldsymbol{r}^*,\boldsymbol{\rho^*})$. If this is not the case, the convergence to the optimal parameters $(\boldsymbol{r}^*,\boldsymbol{\rho^*})$ does not take place and our scheme is unable to satisfy the throughput and awake duration requirements. This makes the results on the finiteness of the optimal parameters $(\boldsymbol{r}^*,\boldsymbol{\rho^*})$ pivotal to our work. To prove this, the authors in \cite{5340575} formulated a different optimization problem where the transmission aggressiveness is assumed to be positive and is taken as the dual variable of an inequality constraint in that new optimization problem. Consequently, they used the Slater's condition in that new OP as an initial step to proceed with the proof of finiteness\color{black}. In our paper, we did not proceed with the same approach. In fact, forcing our new parameter to be positive will result in degradation in terms of energy efficiency and this was therefore omitted. To put this in perspective, suppose that we impose that $\rho_k\geq0$, i.e. $\frac{W_k}{S_k}\geq1$. We can see from this condition that we are forcing the ratio of the waking-up duration to the sleeping duration to be larger than $1$ which is in complete contrast to the energy efficiency aspect of our scheme. This will be further highlighted later on in the simulations section. Therefore, we employed in the following theorem different machinery that helped us to prove the finiteness, baring in mind that we do not impose any conditions on our operational parameters. \color{black}
\begin{theorem} 
	If $\boldsymbol{\lambda} \in \interior(\Lambda)$ and $\boldsymbol{f} \in \interior(\Theta(\boldsymbol{\lambda}))$, then the optimum is attained for a finite $(\boldsymbol{r}^*,\boldsymbol{\rho^*})$
	\label{finite}
\end{theorem}
%To note: A rate $\boldsymbol{\lambda}$ is said to be \emph{strictly} feasible if $\boldsymbol{\lambda} \in int(\mathcal{C})$ or in other words if there exist a neighborhood of $\boldsymbol{\lambda}$ such as every $\boldsymbol{\lambda_I}$ that belongs to this neighborhood is feasible. 
%The case of two interfering links can be found in Figures \ref{feasible} and \ref{waking}. Both $\boldsymbol{\lambda}$ and $\boldsymbol{f}$ that are pointed to verify the required conditions by this theorem.
\begin{IEEEproof} The proof can be found in the Appendix.
\end{IEEEproof}

\section{Implementation Considerations}
\subsection{PDT Assignment}
The PDT parameter assignment can be either \emph{static} (kept constant) or it can be \emph{dynamic} by choosing an appropriate function of the backlog $Q_k$ of each link $k$.\\
\textbf{First case:} $\omega_k$ is assigned to each link and kept constant throughout transmission without change.\\
\textbf{Second case:} Once a large time frame $T_c$ (called \emph{convergence frame}) has elapsed after which the \emph{transmission aggressiveness} and \emph{wake-up  aggressiveness} attain their optimal value, a new constant $\omega_k=g_k(\overline{Q}_k)$ can be assigned to calibrate the power-delay tradeoff based on the average backlog of link $k$ during $T_c$. 
By defining the stochastic backlog process of user k as $Q_k(\tau)$, the average backlog during the time frame $T_c$ can be calculated as follows: 
\begin{equation}
\overline{Q}_k=\frac{\int_{0}^{T_{c}}Q_k(\tau)d\tau}{T_{c}}
\end{equation}
This can be seen as a penalty function for the high queue length, that should be compensated in the next convergence frame. However this chosen function verifies certain properties, the first being that $g_k$ is an increasing function in $\overline{Q}_k$. Other properties include: 
%\begin{equation*}
%g_k(Q_k)\longrightarrow1-\lambda_k \quad \textrm{when} \quad Q_k\longrightarrow+\infty
%\end{equation*}
\begin{equation}
0<g_k(\overline{Q}_k)<1-\lambda_k \quad \forall \overline{Q}_k\geq0
\end{equation}
\subsection{Distributed Gradient Descent}
As previously discussed in Section \Rmnum{3}-B, queuing rate stability of all links in the network, for any arrival rate $\boldsymbol{\lambda} \in \Lambda$, can be achieved by minimizing (\ref{lagrangefunction}) through a distributed gradient algorithm. Moreover, the desired awake duration is achieved. Starting from an arbitrary point $(\boldsymbol{r}_0,\boldsymbol{\rho}_0)$, the updates of both the transmission and waking-up aggressiveness are done each time frame $T_m$ called \emph{update frame}. However, due to the fact that the mixing time of the chain is slow \cite{5340575}, the updates should be done using \emph{sufficiently large} time frames in order to ensure that the chain reaches its stationary distribution. Suppose that links update their aggressiveness at time $t_m$, we define $T_m=t_m-t_{m-1}$ as being the $m$-th update frame. The updates are consequently done as follow:
\begin{equation*}
r_k(m+1)=r_k(m)+\triangle_{1k}(m)(\lambda_k-s_k(m))
\end{equation*}
\begin{equation*}
\rho_k(m+1)=\rho_k(m)+\triangle_{2k}(m)(\lambda_k+\omega_k-\hat{f_k}(m))
\end{equation*}
where $\triangle$ refers to the chosen step size. Both $s_k(m)$\footnote{We suppose that if the queue of link $k$ becomes empty, it sends dummy packets (i.e. the
dummy packets are counted in the throughput computation of $s_k(m)$ in our gradient descent algorithm). This was done to comply with our scheme's assumption that aimed to simplify the Markov chain analysis. This simplification does not alter the stability region of the network (i.e. $s_k=\lambda_k$).} and $\hat{f_k}(m)$ are calculated as follow:
\begin{equation*}
s_k(m)=\frac{\int_{t_m}^{t_{m+1}}a_k(\tau)x_k(\tau)d\tau}{T_{m+1}} \quad \hat{f_k}(m)=\frac{\int_{t_m}^{t_{m+1}}a_k(\tau)d\tau}{T_{m+1}}
\end{equation*}
This process assumes that each link has knowledge of its arrival rate $\lambda_k$. However, if $\lambda_k$ is unknown, it can be approximated using its empirical average. In fact, by defining $V_k(T_a)$ as the total number of packet arrivals for link $k$ in the interval $[0,T_a]$, the arrival rate can be approximated by: 
\begin{equation}
\hat{\lambda}_k=\frac{V_k(T_a)}{T_{a}}
\end{equation}
with $T_a$ being sufficiently large since $\hat{\lambda}_k\rightarrow \lambda_k$ when $T_{a}\rightarrow +\infty$. \\
%A simple note to be taken in consideration, dummy packets are considered to be sent by link $k$ if its buffer was empty to ensure the ideal throughput and awake duration are used during the updates.\\
%The above scheme is exactly a stochastic
%approximation scheme, and is ensured to converge under standard assumptions for
%stochastic approximation.\\
%As we can see from the expression of the \emph{wake-up aggressiveness} and \emph{Transmission aggressiveness},
%the chain depends on the ratio between $\frac{W_k}{S_k}$ and $\frac{R_k}{H_k}$ and not individually on them. For instance, if we fix $S_k$ to a value and update the \emph{wake-up aggressiveness}, it is equivalent to changing just $W_k$ and vice-versa. This means that the node's sleep timer is kept with the same mean but how often the node wakes up depends on the corresponding \emph{wake-up aggressiveness}. We will adopt this approach by fixing both $S_k$ and $H_k$.\\
%Due to the fact that the updates create a fresh new chain, residual effects from the previous chain should be mitigated:
Next, due to the fact that the updates create a fresh new Markov chain, residual effects from the previous chain should be mitigated. 
However, before proceeding, it is essential to point out an interesting aspect of our scheme. Knowing that $\rho_k=\log(W_k/S_k)$, we can see that our scheme depends on the ratio of these two timers' means ($1/S_k$,$1/W_k$) and not individually on one of them. What this means is that we can choose $1/S_k$ as small as desired to let link $k$ sleep quickly if it takes a long time to acquire the channel. This will make $1/W_k$, for a fixed optimal $\boldsymbol{\rho}^*$, to be small as well to keep the same ratio. In other words, link $k$ will wake-up more often (shorter sleeping duration) to contend for the channel (in order to satisfy its throughput requirements) since its sleep timer is chosen to be small. On the other hand, if we choose $1/S_k$ to be high, then we are forcing link $k$ to wait longer in its attempt to transmit data and not to directly transition into SLEEP state. This will result in having a larger $1/W_k$ to maintain the same ratio: in other words, link $k$ will have to wake-up less often (longer sleeping duration) to satisfy its throughput requirement. This analysis has an interesting interpretation: the sleeping duration's mean of link $k$ ($1/W_k$) calibrates itself to how patient we force link $k$ to be in order to still maintain the throughput optimality. In conclusion, although our proposed scheme seems at first in contrast to the line of work in WSN literature since links remain awake when the channel is sensed busy, the calibration property exhibited in our scheme makes it appealing in terms of energy consumption.\\
%However, before proceeding, it is essential to point out that the chain depends on the ratio between $\frac{W_k}{S_k}$ and $\frac{R_k}{H_k}$ and not individually on them. This can be clearly seen from the expression of the \emph{waking-up aggressiveness} and \emph{Transmission aggressiveness}. For instance, we can fix $1/S_k$ to a desired value and therefore when updating the \emph{waking-up aggressiveness}, it is equivalent to changing solely $1/W_k$. By doing so, we are fixing the link's sleep timer mean ( i.e. the mean duration  for which the link is awake and not transmitting before transitioning to SLEEP state) but how often the link wakes up will depend on the corresponding \emph{waking-up aggressiveness}.
%\footnote{For a particular $\boldsymbol{\rho}^*$, if we increase the fixed $1/S_k$, then $1/W_k$ will increase too. This has an interesting interpretation: if $1/S_k$ increases, link $k$ contest more for the channel when it wakes up and therefore its wake-up timer's mean goes up since it is not required to wake-up as often to compensate for its sleeps. On the other extreme, the opposite takes place if we decrease $1/S_k$. During implementation, we are given the freedom to choose between these two extremes, since our scheme will calibrate itself accordingly}. 
In the sequel, we will suppose that both $1/S_k$ and $1/H_k$ are fixed. In this case, to eliminate the residual effects from the previous iteration, the following take place:
\begin{itemize}
\item If the link was awake and transmitting when the update time frame has elapsed, the update of the parameters will not affect it and therefore no further actions are taken
\item If the link was awake but in a back-off stage, a new back-off timer according to the new \emph{transmission aggressiveness} should be chosen 
%A new sleep timer according to the new \emph{wake-up aggressiveness} should be taken. However, if $S_k$ is kept constant as discussed above, then there is no need to get a new sleep timer since the same mean is kept
\item If the link was asleep, a new wake-up timer according to the new \emph{waking-up aggressiveness} should be generated
\end{itemize}
%\begin{rem}
%Knowing that $\rho_k^*=\log(W_k/S_k)$, we can see that our scheme depends on the ratio of these two timers' means ($1/S_k$,$1/W_k$) and not individually on one of them. What this means is that we can choose $1/S_k$ as small as desired to let link $k$ sleep quickly if it takes a long time to acquire the channel. This will make $1/W_k$, for a fixed optimal $\boldsymbol{\rho}^*$, to be small as well to keep the same ratio. In other words, link $k$ will wake up more often (shorter sleeping duration) to contend for the channel (in order to satisfy its throughput requirements) since its sleep timer is chosen to be small. On the other hand, if we choose $1/S_k$ high, then we are telling link $k$ to be patient on the channel and not to directly transition into SLEEP state. This will result in having a larger $1/W_k$ to maintain the same ratio: in other words, link $k$ will have to wake-up less often (longer sleeping duration) to satisfy its throughput requirement.  This analysis has an interesting interpretation: the sleeping duration's mean of link $k$ ($1/W_k$) calibrates itself to how patient we force link $k$ to be in order to still maintain the throughput optimality. In conclusion, although our proposed scheme seems at first in contrasts to the line of work in WSN literature since links remain awake when the channel is sensed busy, the calibration property exhibited in our scheme makes it appealing in terms of energy consumption.
%\end{rem}
\subsection{Contention Resolution}
We have so far focused in this paper on a continuous time CSMA model with no sensing delay. In these idealized settings, collisions are
mathematically impossible, which leads to tractability as a first step of the study and enables us to capture the essence of the scheduling problem without 
being concerned about the contention resolution problem. However, in realistic scenarios, the sensing delay cannot be neglected and therefore the transmission back-off 
timer is chosen as a multiple 
of mini-slots where the duration of the mini-slot $T_{slot}$ is dictated by physical limitations such as propagation delay. In fact, once the wake-up timer of link $k$ ticks, link $k$ 
picks a random back-off timer uniformly distributed from the range $[0,W_k-1]$. In this case, the average back-off timer becomes $T_{slot}\frac{W_k-1}{2}$. We therefore suppose that the contention window $W_k$ of each link $k$ is chosen such that the mean back-off timer $1/R_k$ is preserved which can be calculated from the transmission aggressiveness $r_k$. The back-off timer is decremented whenever the channel is sensed idle for a total mini-slot. In this case, collisions occur when at least two 
interfering links' back-off timers run down at the same time. Clearly, 100\% throughput cannot be achieved in this case. However, as one will see in the sequel, we can achieve throughput close to the maximal allowed by the network. With this model under consideration, one has to compare its performance to the 
continuous counterpart that serves as a \emph{benchmark} (i.e. we cannot achieve better performance than the continuous time version of our scheme). 
For this purpose, we present two distinct approaches:\\
\textbf{\emph{Approach 1}}: Taking into account that $r_k=\log(R_k/H_k)$, we can rewrite the contention window's expression as follows:
\begin{equation}
W_k=\frac{2}{\exp(r_k)\times H_k\times T_{slot}}+1
\label{contentionwindow}
\end{equation}
%Since our scheme gives links the ability to transition into SLEEP state, we define what we will call the \emph{equivalent} contention window as $\overline{W}_k=\frac{W_k}{1-P_{{s}_k}}$ where $P_{{s}_k}$ is the probability for link $k$ to be in SLEEP state. This can be thought as replacing link $k$ by link $k'$ that stays awake continuously and therefore has an average contention window of ${W}_{k'}=\frac{W_k}{1-P_{{s}_k}}$. The motivation behind this comes from the fact that a link that is asleep will not be participating in the contention resolution procedure. In order to have a reasonably low collision probabilities, we seek to lowerbound the \emph{equivalent} contention window of each link (i.e. $\overline{W}_k\geq W_0\:\: \forall k$). In this setting, collisions can be ignored and the performance of the scheme (in both throughput and power consumption) virtually coincides with the continuous time collision-free scenario.
In order to have a reasonably low collision probabilities, we seek to lowerbound the contention window of each link (i.e. $W_k\geq W_0\:\: \forall k$), \emph{conditioned} on link $k$ being awake. In this setting, collisions can be ignored and the performance of the scheme (in both throughput and power consumption) virtually coincides with the continuous time collision-free scenario. 

As it has been previously discussed, links that are assigned low PDT parameter tend to be more aggressive on the channel. Therefore, at first glance, one may think that assigning low PDT parameters to links will lead to an eventual high collisions probability due to the small contention window size as seen in (\ref{contentionwindow}). However, this is not the case due to the dynamic nature of links. In fact, although these links will be aggressive on the channel when they are awake, they spend a decent amount of time in SLEEP state which brings down the number of links in the collision domain. This will drastically reduce the collision probability resulting from the short back-off timers.
 As a matter of fact, it will be shown in the simulations section that we can achieve similar collision performance as the adaptive CSMA while still gaining in terms of power consumption. We can therefore conclude that it is vital to take into account the activity of the links when lowerbounding the contention window. For this purpose, we define what we will call the \emph{equivalent} contention window as $\overline{W}_k=\frac{W_k}{1-P_{{s}_k}}$ where $P_{{s}_k}$ is the probability for link $k$ to be in SLEEP state. This can be thought as replacing link $k$ by link $k'$ that stays awake continuously and therefore has an equivalent contention window of ${W}_{k'}=\frac{W_k}{1-P_{{s}_k}}$. 
%The motivation behind this comes from the fact that a link that is asleep will not be participating in the contention resolution procedure. Therefore, it is allowed to be appropriately more aggressive on the channel when it wakes up since, on average, it will still lead to the same collision probability. 
Hence, we shift our focus to lowerbounding the \emph{equivalent} contention window of each link (i.e. $\overline{W}_k\geq W_0\:\: \forall k$).

%Suppose we want a certain target lower-bound on the contention window for which we have a reasonably low collision probabilities (i.e. $W_k\geq W_0\:\: \forall k$) for which we can approximate our performance with the continous CSMA model. 
To achieve this, first we assume that we are 
given a certain $T_{{slot}_0}$ and mean channel holding time $1/H_k$. Next, to respect the required lower-bound, we are obliged to upperbound the transmission aggressiveness of each link $k$ by a certain value
$r_{max}$. Clearly, by doing so, the maximal throughput achieved by our scheme is reduced since we are limiting how aggressive links can be on the channel. However, as will be shown in the sequel, the performance degradation in realistic scenarios is minor. To see this more clearly, suppose we have $T_{slot}=\SI{9}{\micro s}$ (as adopted in the IEEE 802.11n standard \cite{5307322}), a mean channel holding time of $5$ 
ms and a target contention window lowerbound $W_0=32$. This lowerbound leads to a reasonably low collision probability, assuming that the number of links in a collision domain is not too high (see \cite{840210} for further details). For instance, if links are kept continuously awake, we can see that to achieve the required lowerbound on the equivalent contention window, the transmission aggressiveness $r_k$ of each link $k$ should be upperbounded by $r_{max}=3.5791 \:\: \forall k$. Although this upperbound seems small, it is actually able to achieve a high portion of the maximal throughput of the network. For illustration purposes, we suppose that we are in the case of two interfering links. In this scenario, the capacity region is defined as: 
\begin{equation}
\mathcal{C}=\Big\{\boldsymbol{\lambda}\in\mathbb{R}^{2}_{+} : \lambda_1+\lambda_2\leq 1\Big\}
\end{equation}
By keeping links continuously awake (i.e. by assigning the PDT parameter $\boldsymbol{w}$ to its its maximal value $\boldsymbol{w}_{max}=\boldsymbol{1}-\boldsymbol{\lambda}$ ), the maximum allowed transmission aggressiveness is $r_{max}=3.5791$. When $r_1=r_2=r_{max}=3.5791$, we are able to achieve a total throughput of $s_1+s_2=0.986\approx 1$ which is really close to the maximal throughput of $1$. As for smaller PDT assignments, the following table is presented\footnote{In this framework, we are not limiting the waking-up aggresiveness to a certain value. Therefore, the same performance in terms of power consumption, is to be anticipated. The only performance difference with respect to the benchmark lays in the throughput}:
\begin{center}
\begin{tabular}{|c|c|c|}

 \hline
 Parameter Assignment & $r_{max}$ &Maximal Throughput \\
  \hline
 $\boldsymbol{w}=(\boldsymbol{1}-\boldsymbol{\lambda})$  & $3.579$ & $0.986$ \\
 $\boldsymbol{w}=(\boldsymbol{1}-\boldsymbol{\lambda})/2$  & $3.884$ & $0.980$\\
 $\boldsymbol{w}=(\boldsymbol{1}-\boldsymbol{\lambda})/4$   & $4.091$ & $0.965$\\
 $\boldsymbol{w}=(\boldsymbol{1}-\boldsymbol{\lambda})/8$  & $4.230$ & $0.945$\\
 \hline
\end{tabular}
 \captionof{table}{Maximal Throughput Achieved}
\end{center}
These results suggest that even for extremely low parameter assignments (which translates into a high power gain), we are still able to achieve decent performance, even when collisions are taken into account\footnote{For instance, in the last table entry, $\boldsymbol{\lambda}=(0.4725,0.4725)$ and the parameter assignment is $\boldsymbol{w}=(0.0659,0.0659)$ which is considerably low and therefore would achieve high reduction in power consumption}. Moreover, we can achieve even higher throughput by increasing the mean channel holding time of each link $k$ since, for a given $W_0$, it will make $r_{max}$ increase.\\
\textbf{\emph{Approach 2}}: In this approach, we recall that in order to achieve throughputs under slotted CSMA algorithms close to those 
obtained under the continuous-time CSMA algorithm, it is sufficient to keep the collision duration $T_{collision}$ small with respect to the channel holding time (as presented and tested in \cite{5336355}). To accomplish this in our framework, we can adopt a variant of our scheme that takes into account the discrete nature of the back-off stage and approaches the optimal 
performance of the CSMA scheme. When the back-off timer of link $k$ runs out, the link probes the channel with a small signaling message of duration $\delta$ similar 
to the RTS/CTS mechanism adopted in IEEE 802.11. In this case, when a collision takes place, only these small signaling messages collide and therefore the collision 
duration is limited to $\delta$. By having an average transmission time $1/H_k$ large compared to $\delta$, the throughputs achieved by each link are close to those of the 
continuous time counterpart. Of course, the incorporation of the RTS/CTS mechanism increases the overhead of successful transmissions. However, with this overhead 
being small compared to the overall transmission time, this mechanism comes at a fairly small penalty in terms of overall throughput $s_k\: \forall k$ and awake duration $f_k\: \forall k$ (and therefore power consumption) and the performance would 
virtually coincide with thus of the continuous time counterpart.
\section{Simulations}
\begin{figure}[!ht]
    \centering
    \includegraphics[scale=0.41]{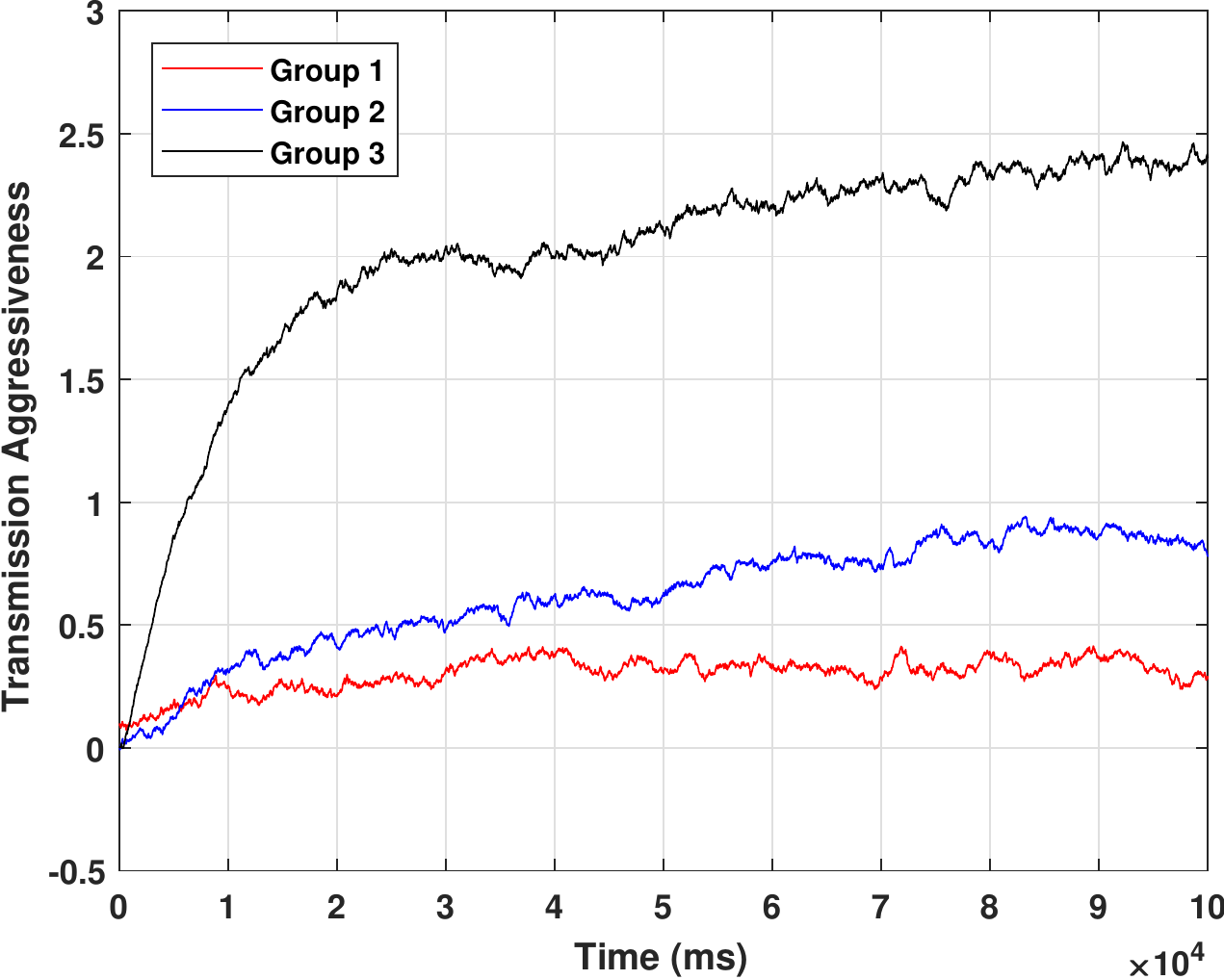}
    \caption{Evolution of the transmission aggressiveness}
    \label{fig:Aggresiveness1}
    \vspace{-8pt}
\end{figure}

\begin{figure}[!ht]
    \centering
    \includegraphics[scale=0.41]{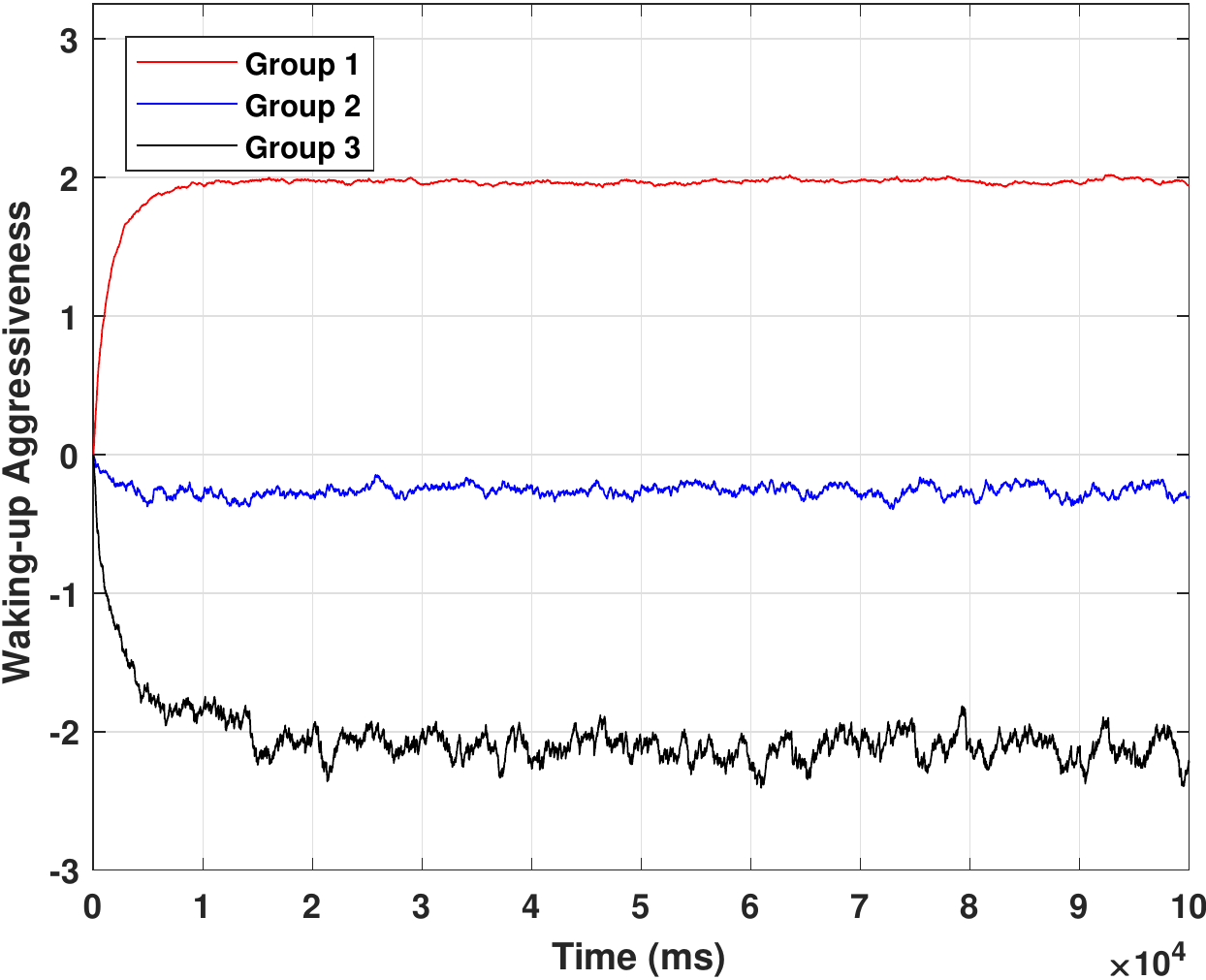}
    \caption{Evolution of the waking-up aggressiveness}
    \label{fig:Aggresiveness2}
    \vspace{-8pt}
\end{figure}

The goal of these simulations is to corroborate the theoretical results in terms of parameters convergence and highlight the power gain experienced by the links using our proposed scheme in comparison to the adaptive CSMA counterpart. Moreover, the provided simulations put into perspective how the IEEE 802.11 protocol fails as the load on the network increases. Before proceeding to the simulations, we first present the power model taken into consideration:
\begin{itemize}
\item  When the link is in a \emph{SLEEP} state, the link consumes $P_z$
\item When the link is transmitting, it consumes $P_t$
\item When the link is sensing the medium, it consumes $P_s$. The link essentially receives radio signals and after signal processing, it takes a decision whether the medium is to be considered busy or not. We neglect the processing power consumption and assume that the power consumption in this case is simply the power required to receive these radio signals (i.e. $P_s=P_r$)
\end{itemize}
%We will show that even with this assumption that gives the adaptive CSMA an edge, since our goal is to minimize the power induced by the medium sensing, we will still achieve a decent power gain. 
As for the numerical values, we will adopt the values of CC1101 in Table \ref{tablepower}, a low power RF transceiver proposed by the industry \cite{CD4007}. It is worth mentioning that using the parameters of other low power transceivers yields similar relative results.
\begin{center}
\begin{tabular}{|c|c|}
 \hline
 \multicolumn{2}{|c|}{Power Model} \\
 \hline
 Power Parameter & Value \\
  \hline
 $P_z$  & \SI{1.5}{\micro\watt} \\
 $P_t$   & \SI{73}{\milli\watt} \\
 $P_r$   & \SI{45}{\milli\watt} \\
 \hline
\end{tabular}
 \captionof{table}{Power model values}
 \label{tablepower}
 \vspace{-4pt}
\end{center}
We assume that the channel mean holding time is 1 ms, and as mentioned before, we are going to assume that the sleeping timer's mean is kept constant throughout the simulations, which is fixed to 1 ms in the sequel. We consider a realistic heterogeneous case where several groups of links exist in the network with each group having its own desired PDT. The number of groups is chosen as $3$ with $4$ links in each group:\\
-\emph{Group 1:} This group is made of links that are delay sensitive 
but can tolerate a high power consumption\\
-\emph{Group 2:} This group is made of links that fall between the two extremes, 
they require a moderate power consumption without introducing a lot of delay\\
-\emph{Group 3:} This group is made of links that can tolerate long delays however they are extremely power limited\\ 
We consider that the arrival for each link is $\lambda_k=0.077 \:\: \forall k$. $\omega_1=0.8$, $\omega_2=0.4$ and $\omega_3=0.1$ are assigned for Groups 1,2 and 
3 respectively. The simulations are run for $100\:s$ with a fixed update time 
frame $T_m=10\:ms$ and with the same step size $\triangle=0.1$ for all updates.\\
\textbf{Parameters Convergence:} 
%For the adaptive CSMA, the minimization of the 
%convex objective function leads to a theoretical \emph{Transmission 
%Aggressiveness} vector $\boldsymbol{r^*}=(0.154,0.154)$. 
By solving our optimization problem previously stated in our theoretical analysis (see Section \Rmnum{3}-B), the optimal aggressiveness parameters can be found in the following table\footnote{Links within the same group share the same optimal aggressiveness parameters}\footnote{If we have imposed positivity of $\boldsymbol{\rho}$, links of group $2$ and $3$ will be forced to stay awake longer and the performance in terms of energy efficiency will be hugely degraded.}:\\
\begin{center}
\begin{tabular}{|c|c|c|c|}
 \hline
 \multicolumn{4}{|c|}{Optimization Problem Solutions} \\
 \hline
 \multicolumn{3}{|c|}{Proposed Scheme} & Adaptive CSMA \\
 \hline
 Group Index & $r^*$ & $\rho^*$  & $r^*$\\
  \hline
 $1$  & $0.1561$ & $1.8724$ & $0.014$\\
 $2$   & $0.8492$ & $-0.2681$ & $0.014$ \\
 $3$   & $2.2355$ & $-2.1078$ & $0.014$ \\
 \hline

\end{tabular}
 \captionof{table}{Optimal Aggressiveness}
\end{center}
It can be 
seen from these values that by adapting our scheme, links are more aggressive on the channel when they are awake compared to the adaptive CSMA (which translates into smaller back-off timers). Moreover, links with low power-delay tradeoff parameter (i.e. links that spend more time in SLEEP state) are more aggressive on the 
channel when they wake-up. The extra aggressiveness can be thought to be a \emph{compensation} for the time spent in SLEEP state. In fact, to be able to maintain their ability to withstand their arrival rate, links that spend a high amount of time in SLEEP state have to be aggressive on the channel to rapidly capture it. As demonstrated in both Fig. \ref{fig:Aggresiveness1} and \ref{fig:Aggresiveness2}, the optimum parameters are achieved distributively by each link, simply by monitoring: \begin{enumerate}
\item the amount of time it spends being awake
\item the amount of time it spends transmitting\footnote{For ease of presentation, we have only shown the parameters convergence for an arbitrary link in each group while noting that links in each group experienced the same behavior}
\end{enumerate}
\begin{figure*}[ht]
\centering
\begin{subfigure}{0.33\textwidth}
  \centering
  \includegraphics[width=.9\linewidth]{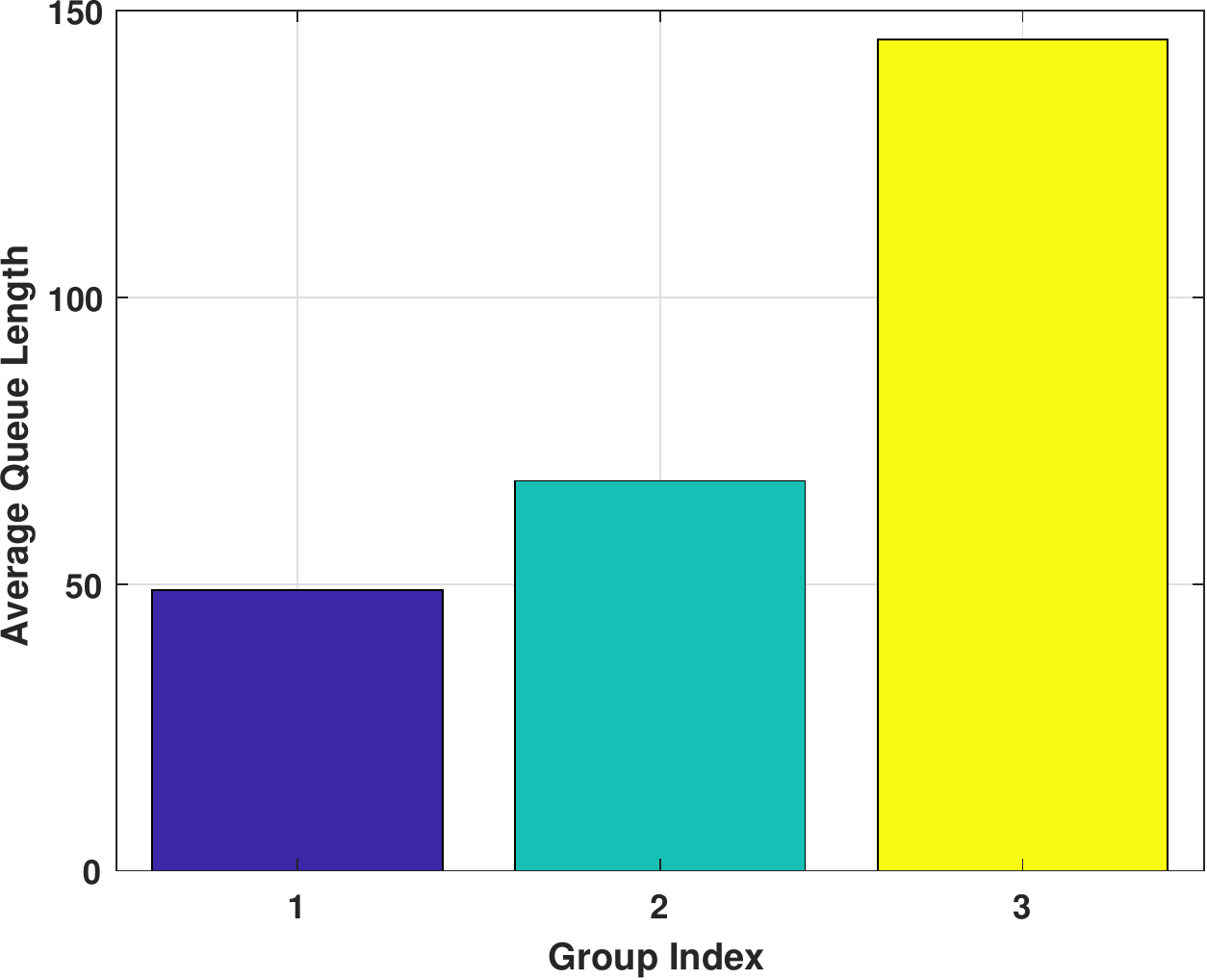}
  \caption{Average queue length of each group}
    \label{fig:queue12}
\end{subfigure}%    
\begin{subfigure}{0.33\textwidth}
\centering
  \includegraphics[width=.9\linewidth]{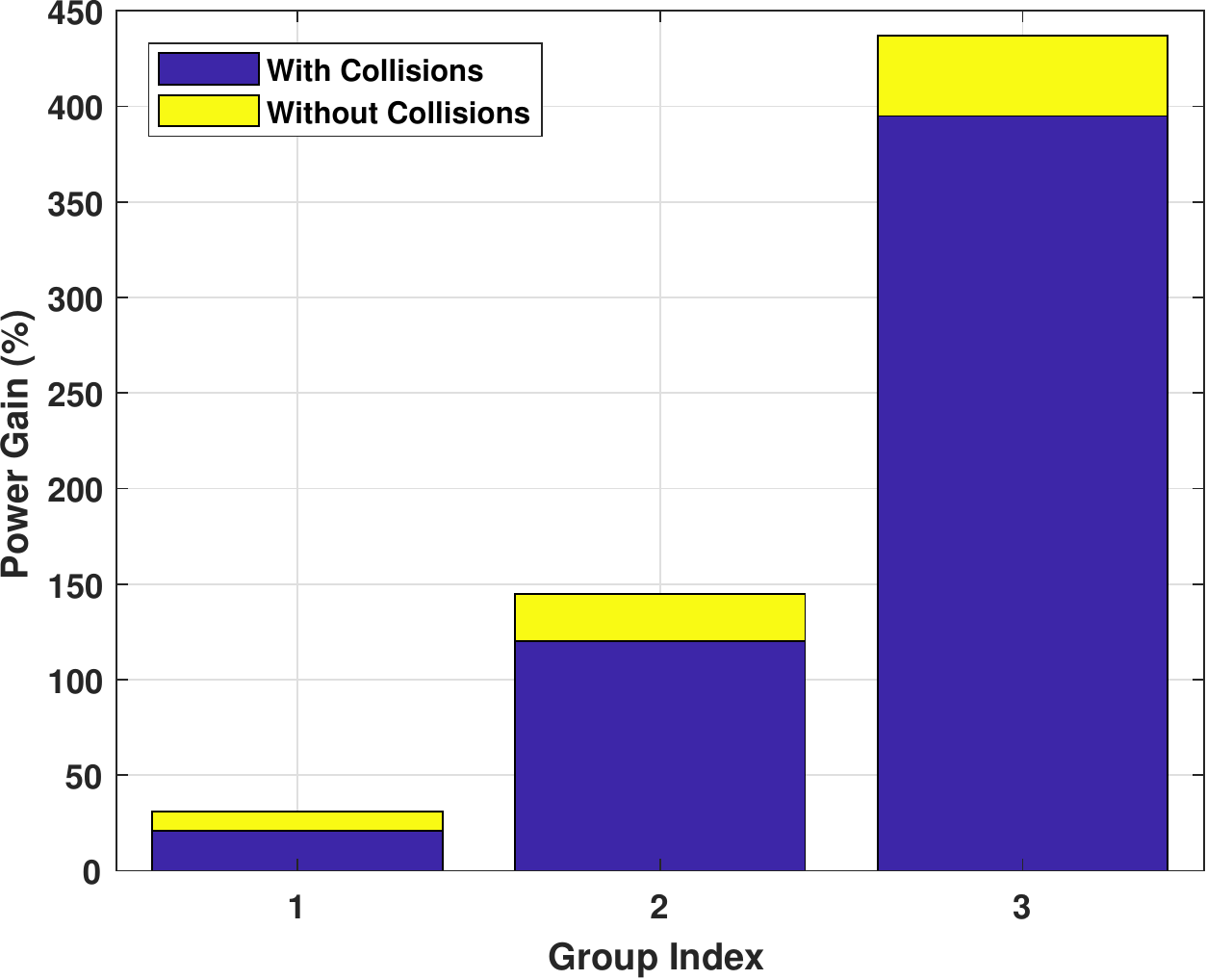}
  \caption{Comparison with the adaptive CSMA in terms of energy}
\label{wifilowenergy}
\end{subfigure}%
\begin{subfigure}{0.33\textwidth}
  \centering
  \includegraphics[width=.9\linewidth]{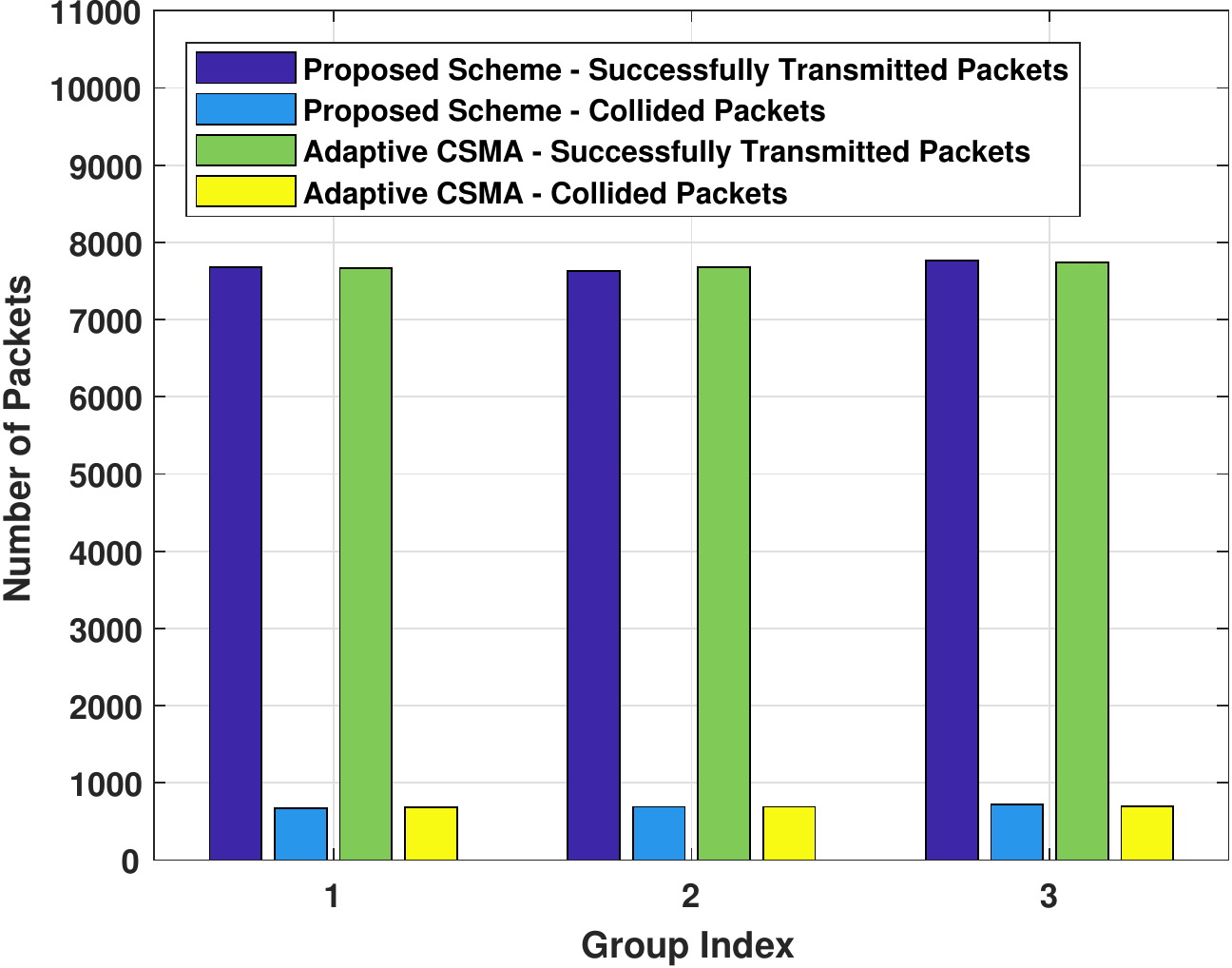}
  \caption{Comparison in terms of successfully transmitted and collided packets}
\label{wifilowthroughput}
\end{subfigure}%

%\begin{subfigure}{0.25\textwidth}
%  \centering
%  \includegraphics[width=.99\linewidth]{wifihighthroughput}
%  \caption{\color{blue} Comparison in terms of successfully transmitted and collided packets - High load}
%\label{wifihighthroughput}
%\end{subfigure}

\caption{Simulations Results \color{black}}

\vspace{-20pt}
\end{figure*}
\noindent\textbf{Energy Efficiency:} In order to best demonstrate the difference in performance between our scheme and the standard adaptive CSMA, we let the initial aggressiveness 
parameters to be close to their optimal values for both schemes.

The results are presented in both Fig.\ref{fig:queue12} and Fig.\ref
{wifilowenergy} (without collisions) where the power is evaluated per successfully transmitted packet. As expected, Group 3 experienced the highest queue length but benefited from a 
high power gain (recall that links in Group 3 are delay tolerant). Group 1 on the other 
extreme had the complete opposite results while Group 2 laid in between these 
two cases. Fig.\ref{wifilowenergy} shows the gain in terms of power of our scheme 
as compared to the adaptive CSMA. Our results show clearly that our scheme 
can handle multiple services at the same time by simply adjusting the 
PDT parameter. For some IoT applications (with delay tolerance), our scheme 
provides a huge power gain.

\begin{figure}[!ht]
    \centering
    \includegraphics[scale=0.48]{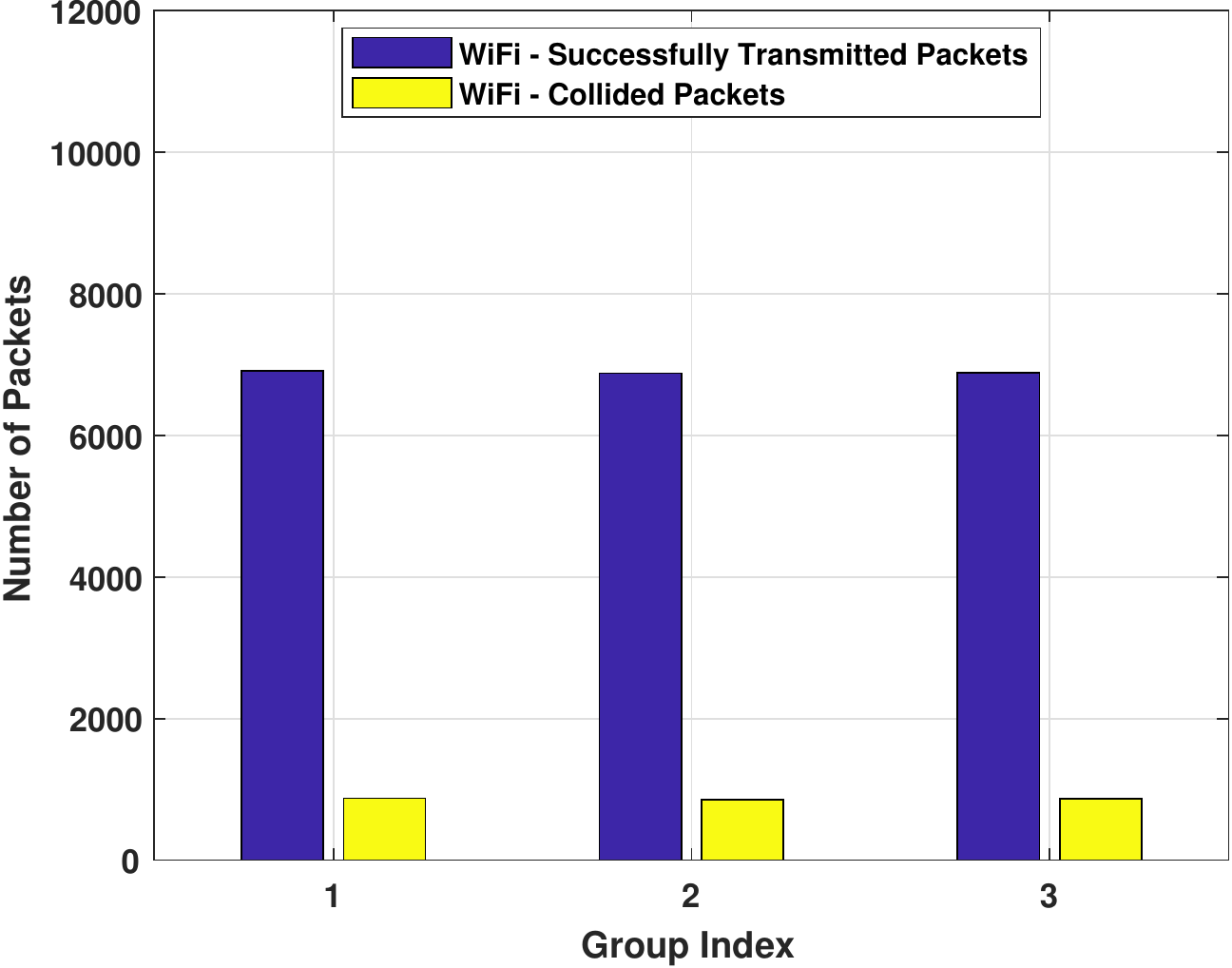}
    \caption{IEEE 802.11 performance in terms of successfully transmitted and collided packets}
    \label{wifiperf}
    \vspace{-18pt}
\end{figure}

\noindent\textbf{The effect of collisions:} In this scenario, we relax the assumption of zero sensing delay. Therefore, collisions cannot be ignored any longer and a performance degradation with respect to the continuous time counterpart is to be expected. The mini-slot is chosen to be $T_{slot}=\SI{9}{\micro s}$ (as adopted in the IEEE 802.11n standard \cite{5307322}). While taking into account the collisions, the following performance indicators are investigated:
\begin{enumerate}
\item Total number of successfully transmitted packets
\item Power gain with respect to the theoretical (collision-free) continuous time adaptive CSMA
\end{enumerate}
We can see from the results in Fig. \ref{wifilowenergy} and \ref{wifilowthroughput} that although the arrival rate $\boldsymbol{\lambda}$ is close to the capacity region's boundary ($\sum_{k=1}^{12}\lambda_k=0.924$ close to the maximal throughput of $1$) and the collision domain's density is high ($12$ links in a single collision domain), we are still able to achieve performance close to the collision-free performance
%\footnote{It is worth mentioning that in this case, the collided packets are not counted in the transmission aggressiveness gradient updates computations} 
(a small degradation in terms of power gain is witnessed, which is natural due to the power lost on collided packets). Moreover, although links are aggressive on the channel using our scheme (especially those with low PDT parameters), we can see that the ratio of the lost packets to the total transmitted packets (i.e. the collision probability) is similar to the adaptive CSMA counterpart. This comes from the dynamic nature of the activity of links. In other words, although these links are aggressive on the channel, they spend a decent amount in SLEEP state which reduces the probability of collisions with other links. In summary, the performance of our scheme is close to the theoretical performance due to several factors:
\begin{enumerate}
\item The network consists of several groups with different degree of aggressiveness on the channel (e.g. some links are more aggressive on the channel than the others)
\item The dynamic nature of the activity of links: although several links are aggressive, they may be in SLEEP state when other links are contending for the channel
\end{enumerate}
\noindent\textbf{Comparison with IEEE 802.11:} To further highlight the advantages of our proposed scheme, we compare it with IEEE 802.11 (Wi-Fi) as well. The configuration settings of IEEE 802.11 were set as follow:
\begin{itemize}
\item Binary exponential backoff is used with a maximum of $10$ multiplications (i.e. CW$_{max}$=1024$\times$CW$_0$)
\item The contention window CW$_0$ is set based on the work of Bianchi to achieve the highest possible throughput \cite{840210}
\end{itemize} 
We can see in Fig. \ref{wifiperf} that IEEE 802.11 is able to achieve a maximum throughput of around $s_k=0.068\:\: \forall k$ (i.e the maximal total throughput is $\sum_{k=1}^{12}s_k=0.816$). This means that when $\lambda_k=0.077 \:\: \forall k$, the queue length of each link will grow indefinitely when IEEE 802.11 is used. This comes from the fact that IEEE 802.11 suffers from being throughput suboptimal \cite{5340575}. On the other hand, one can clearly see in Fig. \ref{wifilowthroughput} that both the adaptive CSMA and our proposed scheme satisfy the requirements in terms of throughput $s_k\thickapprox \lambda_k$ (each link sent $N$ successfully packets where $N\thickapprox0.077\times10^5=7700$). With our proposed scheme being throughput optimal, one can therefore expect an increase of the maximal possible throughput of around $20\%$ with respect to IEEE 802.11. As for the power consumption, IEEE 802.11 experienced a $15\%$ loss of power with respect to the adaptive CSMA. To conclude, our proposed scheme is able to satisfy the throughput requirements of links in high load environments due to its throughput optimality while providing huge power gain with respect to the adaptive CSMA.

%\begin{figure}[!ht]
%    \centering
%    \includegraphics[scale=0.45]{wifihighenergy}
%    \caption{\color{blue} Comparison with the adaptive CSMA in terms of power consumption - High Load}
%    \label{wifihighenergy}
%    \vspace{-15pt}
%\end{figure}
\color{black}
%
%\begin{figure*}[!ht]
%\centering
%\begin{subfigure}{0.5\textwidth}
%  \centering
%  \includegraphics[width=.4\linewidth]{wifihighthroughput}
%  \caption{Comparison of sum spectral efficiency vs. SNR}
%  \label{rate}
%\end{subfigure}%
%\begin{subfigure}{0.5\textwidth}
%  \centering
%  \includegraphics[width=.4\linewidth]{wifihighenergy}
%  \caption{Comparison of Jain's fairness index vs. SNR}
%  \label{fairness}
%\end{subfigure}%
%\caption{Scheduling Schemes Comparison}
%\end{figure*}
\section{Conclusion}
In this paper, we have introduced a new MAC scheme that belongs to the CSMA family. In this scheme, with the aim of reducing power consumption, each link is allowed to 
transition between AWAKE and SLEEP states. By controlling operational parameters such as back-off and sleeping timers with the aim of optimizing a certain 
objective function, we were able to show that our scheme is throughput optimal. The convergence of the parameters to their optimal values has been proven to be 
completely distributed without any message passing. The theoretical analysis resulted in the birth of a parameter which had the interpretation of being a power-delay 
tradeoff. This parameter is assigned to each link depending on the application concerned. Implementation considerations were provided and the simulations conformed 
with the theoretical results and showed the performance advantages in terms of power gain with respect to the adaptive CSMA. In the future, the authors will focus
on the study of the convergence of the proposed CSMA scheme along with a careful investigation of the convergence speed.\color{black}
\section{Acknowledgments}
The authors would like to express their gratitude to the Associate Editor and the reviewers for their many constructive
comments and suggestions to improve the quality of the paper.
\bibliographystyle{IEEEtran}
\bibliography{trialout}
\appendix
\section*{Proof of Theorem 2}
We recall that our original optimization problem is to minimize the \textit{Kullback-Leibler} divergence between the joint probabiblity distribution $\boldsymbol{p}$ and our Markov chain's stationary distribution. Knowing that $D(\boldsymbol{p}\|\pi(\boldsymbol{r},\boldsymbol{\rho}))\geqslant0$, we can conclude: 
\begin{equation*}
\begin{aligned}
& \underset{\boldsymbol{r},\boldsymbol{\rho}}{\inf}
& & D(\boldsymbol{p}\|\pi(\boldsymbol{r},\boldsymbol{\rho})) \: \: \textrm{exists}
\end{aligned}
\end{equation*}
Now what remains is to show that the minimum is attained by a finite $(\boldsymbol{r^*},\boldsymbol{\rho^*})$. The motivation behind this theorem's assumptions are the following:
\begin{itemize}
\item if $f_k=\lambda_k$: the only way to achieve this is for link $k$ to wake-up and instantly acquire the channel. Consequently for this case, the value of $r_k$ should 
tend to $+\infty$
\item if $f_k=1$: the only way to achieve this is for link $k$ to continuously stay awake. Consequently for this case, the value of $\rho_k$ should tend to $+\infty$
\item As for the strict feasibility\footnote{A rate $\boldsymbol{\lambda}$ is said to be \emph{strictly} feasible if $\boldsymbol{\lambda} \in \interior(\Lambda)$ or in other words if there exist a neighborhood of 
$\boldsymbol{\lambda}$ such that every $\boldsymbol{\lambda_I}$ that belongs to this neighborhood is feasible} of $\boldsymbol{\lambda}$: when the arrival rate belongs to the boundary of the capacity region, links become extremely aggressive since the network hits its limit and consequently $r_k \:\: \forall k$ should 
tend to $+\infty$. 
\end{itemize}
We provide a rigorous proof to our claim in the sequel. 
The proof is divided into two sections: first, a lemma is provided to show the equivalence between the assumptions of the theorem and the strict positivity of the joint probability distribution $\boldsymbol{p}$ where:
$$\begin{cases}\lambda_k=\sum\limits_{j=1}^{2^K}\sum\limits_{i=1}^{|I_j|}p_{ij}a_k^jx_k^i \\
f_k=\sum\limits_{j=1}^{2^K}\sum\limits_{i=1}^{|I_j|}p_{ij}a_k^j\\
\end{cases}$$
The results of this Lemma will be used to proceed with the proof by contradiction and making use of a fundamental mathematical theorem on bounded series: the ``Bolzano\textendash Weierstrass Theorem".
In fact, the proof by contradiction assumes that $p_{ij}>0 \:\: \forall(i,j)$ which makes the results of this Lemma pivotal to the rest of the proof.
\begin{lemma}
The joint probability distribution $\boldsymbol{p}$ satisfies $p_{ij}>0\:\forall i,j$ if and only if $$\begin{cases} \lambda_k>0\: \:\forall k \\
\boldsymbol{\lambda}\: \: \textrm{is strictly feasible} \\
\lambda_k <f_k<1 \:\:\forall k
\end{cases}$$
\label{positivity}
\end{lemma}
\begin{IEEEproof} The proof of this Lemma is divided into two parts where, in each part, one of the required implications is proven.\\
\emph{\textbf{Part 1:}} We start by proving that if $p_{ij}>0\:\forall i,j$ then:
$$\begin{cases} \lambda_k>0\: \:\forall k \\
\boldsymbol{\lambda}\: \: \textrm{is strictly feasible} \\
\lambda_k <f_k<1 \:\:\forall k
\end{cases}$$\\
- $\boldsymbol{f}<\boldsymbol{1}:$ the state $S=(\boldsymbol{0},\boldsymbol{0})$ has a non-null probability then no link stay awake all the time and therefore we have $f_k<1 \:\:\forall k$\\
- $\boldsymbol{f}>\boldsymbol{\lambda}:$ states where links are awake and not transmitting have non zero probability i.e. links enter in back-off stages and therefore spend more time awake than transmitting\\
- $\boldsymbol{\lambda}>\boldsymbol{0} :$ it is straightforward since each link has an arrival rate of at least $p_k=(a^j=\boldsymbol{e_k},x^i=\boldsymbol{e_k})>0$\\
- \emph{Strict feasibility} : we will have to prove that there is exist a neighborhood of $\boldsymbol{\lambda}$ such that each arrival rate inside of it is feasible. Let $p_0=P((\boldsymbol{1},\boldsymbol{0}))>0$ the portion of time users are all awake and none of them is transmitting and $p_k=P((\boldsymbol{1},\boldsymbol{e_k}))>0$ where $\boldsymbol{e_k}$ refers to the $k^{th}$ canonical vector, we define  $\epsilon=min\{p_0/K,\displaystyle{\min_{k} p_k}\}>0$. We can satisfy any arrival rate $\boldsymbol{\lambda'}$ such that:
\begin{equation*}
|\lambda_k'-\lambda_k|\leqslant \epsilon \:\: \forall k
\end{equation*}
We can do that by simply constructing the following distribution $\boldsymbol{p'}$ such that:
$$\begin{cases} p_0'=p_0-\sum\limits_{k=1}^{K}(\lambda_k'-\lambda_k) \\
p_k'=p_k+\lambda_k'-\lambda_k \:\: \forall k \\
p'=p\: \: \textrm{for all other states}
\end{cases}$$\\
To note that in this case, $f_k'=f_k \:\: \forall k$.\\
\emph{\textbf{Part 2:}} Throughout this section, a visualization of the proof in the case of two interfering links is presented to clarify the explanation. In this part, we suppose that $\boldsymbol{\lambda}$ and $\boldsymbol{f}$ verify: 
$$\begin{cases} \lambda_k>0\: \:\forall k \\
\boldsymbol{\lambda}\: \: \textrm{is strictly feasible} \\
\lambda_k <f_k<1 \:\:\forall k
\end{cases}$$\\
We will try to construct a joint probability distribution $\boldsymbol{p}$ that satisfies $p_{ij}>0\:\forall i,j$ such that:
$$\begin{cases}\lambda_k=\sum\limits_{j=1}^{2^K}\sum\limits_{i=1}^{|I_j|}p_{ij}a_k^jx_k^i \\
f_k=\sum\limits_{j=1}^{2^K}\sum\limits_{i=1}^{|I_j|}p_{ij}a_k^j\\
\end{cases}$$\\
We start by choosing an arbitrary probability distribution $\boldsymbol{p_A>0}$, then four cases can occur:\\
(a): the randomly constructed $\boldsymbol{p_A}$ verifies the requirements $\boldsymbol{\lambda_A}=\boldsymbol{\lambda}$ and $\boldsymbol{f_A}=\boldsymbol{f}$ then we choose $\boldsymbol{p}=\boldsymbol{p_A}$\\
(b): $\boldsymbol{p_A}$ verifies $\boldsymbol{\lambda_A}=\boldsymbol{\lambda}$ but $\boldsymbol{f_A}\neq \boldsymbol{f}$. Since both $\boldsymbol{f_A},\boldsymbol{f}>\boldsymbol{\lambda}$ (for $\boldsymbol{f_A}$, we recall part 1 since $\boldsymbol{p_A>0}$) hence they lay in the same region and we can find $d>0$ such that:
\begin{equation*}
\boldsymbol{f_{B}}=\boldsymbol{f}+d(\boldsymbol{f}-\boldsymbol{f_{A}})
\end{equation*}
Therefore we can find $\boldsymbol{p_B}\geqslant 0$ such that $\boldsymbol{\lambda_B}=\boldsymbol{\lambda}$. We can then construct in this case $\boldsymbol{p}$ in the following way: 
\begin{equation*}
p_{ij}=\theta p_{A_{ij}}+(1-\theta)p_{B_{ij}}
\end{equation*}
where $\theta=\frac{d}{1+d}>0$. We know that $p_{A_{ij}}>0$ and $p_{B_{ij}}\geqslant0$, therefore $p_{ij}>0$.\\
%\begin{figure}[!ht]
%\centering
%\includegraphics[width=.8\linewidth]{text7677.png}
%\setlength{\belowcaptionskip}{-10pt}
%\caption{Case (b) capacity region}
%\label{case b capacity}
%\end{figure}
%\begin{figure}[!ht]
%\centering
%\includegraphics[width=.82\linewidth]{text5970.png}
%\setlength{\belowcaptionskip}{-10pt}
%\caption{Case (b) awake region}
%\label{case b awake}
%\end{figure}\\
(c): $\boldsymbol{p_A}$ verifies the requirement for $\boldsymbol{f_A}=\boldsymbol{f}$ but $\boldsymbol{\lambda_A}\neq \boldsymbol{\lambda}$. It is tricky here since the region of $\boldsymbol{f}$ depends on $\boldsymbol{\lambda}$. In other words, $\boldsymbol{f_A}$ and $\boldsymbol{f}$ do not lay in the same region. However since $\boldsymbol{\lambda}<\boldsymbol{f}$, we can always find a neighborhood of $\boldsymbol{\lambda}$ such that for each $\boldsymbol{\lambda_I}$ in it, $\boldsymbol{\lambda_I}$ verifies $\boldsymbol{\lambda_I}<\boldsymbol{f}$. 
%This neighborhood is marked as a red disk in Figure \ref{casecanalysis}. 
Now we take the intersections between this neighborhood and all the affine combinations between $\boldsymbol{\lambda}$ and $\boldsymbol{\lambda_A}$, then we can state that there is exist $\boldsymbol{\lambda_B}$ in this intersection (hence verifies $\boldsymbol{\lambda_B}<\boldsymbol{f}$) such that $\exists d>0$ in a way that:
\begin{equation*}
\boldsymbol{\lambda_{B}}=\boldsymbol{\lambda}+d(\boldsymbol{\lambda}-\boldsymbol{\lambda_{A}})
\end{equation*}
Let $\boldsymbol{p_B}\geqslant 0$ its corresponding probability distribution with $\boldsymbol{f_B}=\boldsymbol{f}$ which is possible since $\boldsymbol{\lambda_B}<\boldsymbol{f}$. Then we can conclude that we can write $\boldsymbol{p}$ in the following way:\\
\begin{equation*}
p_{ij}=\theta p_{A_{ij}}+(1-\theta)p_{B_{ij}}
\end{equation*}
where $\theta=\frac{d}{1+d}>0$. We know that $p_{A_{ij}}>0$ and $p_{B_{ij}}\geqslant0$, therefore $p_{ij}>0$.
%\begin{figure}[!ht]
%\centering
%\includegraphics[width=.8\linewidth]{"CASE 2".png}
%\setlength{\belowcaptionskip}{-10pt}
%\caption{Case (c) capacity region}
%\label{case c capacity}
%\end{figure}
%\begin{figure}[!ht]
%\centering
%\includegraphics[width=.82\linewidth]{path4925.png}
%\setlength{\belowcaptionskip}{-10pt}
%\caption{Case (c) awake region}
%\label{casec awake}
%\end{figure}\\
%\begin{figure}[!ht]
%\centering
%\includegraphics[width=.71\linewidth]{text7677-6.png}
%\setlength{\belowcaptionskip}{-10pt}
%\caption{Visualization of case (c) analysis}
%\label{casecanalysis}
%\end{figure}\\
(d): in this case we have $\boldsymbol{f_A}\neq \boldsymbol{f}$ and $\boldsymbol{\lambda_A}\neq \boldsymbol{\lambda}$. This might seems like the hardest case but it is merely a combination of both cases (b) and (c). Since the priority is to coincide the regions of $\boldsymbol{f_A}$ and $\boldsymbol{f}$, we start by using the same analysis as case (c). By doing so, we have:
\begin{equation*}
\boldsymbol{\lambda_{B}}=\boldsymbol{\lambda}+d_1(\boldsymbol{\lambda}-\boldsymbol{\lambda_{A}})
\end{equation*}
and let $\boldsymbol{p_B}\geqslant 0$ be its corresponding probability distribution with $\boldsymbol{f_B}=\boldsymbol{f}$. We construct the probability distribution $\boldsymbol{p_C}$ as follows:
\begin{equation*}
p_{C_{ij}}=\theta_1 p_{A_{ij}}+(1-\theta_1)p_{B_{ij}}
\end{equation*}
where $\theta_1=\frac{d_1}{1+d_1}>0$. We know that $p_{A_{ij}}>0$ and $p_{B_{ij}}\geqslant0$, we can conclude that $p_{C_{ij}}>0$. We now have $\boldsymbol{\lambda_C}=\boldsymbol{\lambda}$ but $\boldsymbol{f_C}\neq \boldsymbol{f}$. We are back to case (2) and therefore, we can find $d_2>0$ such that:
\begin{equation*}
\boldsymbol{f_{D}}=\boldsymbol{f}+d_2(\boldsymbol{f}-\boldsymbol{f_{C}})
\end{equation*}
and $\boldsymbol{\lambda_D}=\boldsymbol{\lambda}$. We can then construct in this case $\boldsymbol{p}$:
\begin{equation*}
p_{ij}=\theta_2 p_{C_{ij}}+(1-\theta_2)p_{D_{ij}}
\end{equation*}
where $\theta_2=\frac{d_2}{1+d_2}>0$. Knowing that $p_{C_{ij}}>0$ and $p_{D_{ij}}\geqslant0$, we have $p_{ij}>0$ which concludes the proof of the Lemma
\end{IEEEproof}
The next step is to use the results of Lemma. \ref{positivity} to prove that our optimum is attained for a finite $(\boldsymbol{r^*},\boldsymbol{\rho^*})$. As proven in Lemma. \ref{positivity}, our theorem's assumptions are equivalent to supposing that $p_{ij}>0\:\forall (i,j)$. Taking that into consideration, we consider in the following that $p_{ij}>0\:\forall i,j$. First, we recall our optimization problem:
\begin{equation*}
\begin{aligned}
& \underset{\boldsymbol{r},\boldsymbol{\rho}}{\text{minimize}}
& & F(\boldsymbol{r},\boldsymbol{\rho})=-\sum_{j=1}^{2^K}\sum_{i=1}^{|I_j|}p_{ij}\log(\pi(a^j,x^i;\boldsymbol{r},\boldsymbol{\rho}))
\end{aligned}
\end{equation*}
We have previously proven in Section \Rmnum{3}-B that this minimization is indeed a convex optimization problem after appropriate transformations. Therefore, we either have one of the following cases:\\
(a): a finite minimizer $(\boldsymbol{r^*},\boldsymbol{\rho^*})$ exists and it is unique due to the convexity of our optimization problem\\
(b): there exist a sequence $(\boldsymbol{r}^n)_n$ such that
$F((\boldsymbol{r}^n)_n,\boldsymbol{\rho^*}) \overset{n\rightarrow+\infty}{\longrightarrow} F^*$ and $||\boldsymbol{r}^n|| \overset{n\rightarrow+\infty}{\longrightarrow} +\infty$ \\
(c): there exist a sequence $(\boldsymbol{\rho}^n)_n$ such that
$F(\boldsymbol{r^*},(\boldsymbol{\rho}^n)_n) \overset{n\rightarrow+\infty}{\longrightarrow} F^*$ and $||\boldsymbol{\rho}^n|| \overset{n\rightarrow+\infty}{\longrightarrow} +\infty$ \\
(d): there exist two sequences $(\boldsymbol{r}^n)_n$ and $(\boldsymbol{\rho}^m)_m$ such that $F((\boldsymbol{r}^n)_n,(\boldsymbol{\rho}^m)_m)\overset{n,m\rightarrow+\infty}{\longrightarrow}F^*$ while the norm of these two vectors $||\boldsymbol{r}^n||,||\boldsymbol{\rho}^m|| \overset{n,m\rightarrow+\infty}{\longrightarrow} +\infty$.\\
The main idea revolves around proving that case (a) is the only possible outcome of the optimization problem. For this purpose, we recall a fundamental mathematical theorem on bounded sequences. Before stating the theorem, we call to mind the notion of accumulation point. Let $(\boldsymbol{a}_n)$ be a sequence of real vectors, the vector $\boldsymbol{L}$ is said to be an \emph{accumulation} point of $(\boldsymbol{a}_n)$ if there exists a subsequence $(\boldsymbol{a}_{n_k})$ that converges to $\boldsymbol{L}$. In other words:
\begin{equation*}
\forall\epsilon>0, \exists K \in \mathbb{N} \: \textrm{such that if} \: k\geqslant K \: \textrm{then} \: ||\boldsymbol{a}_{n_k}-L||\leqslant\epsilon
\end{equation*}
\begin{theorem}[Bolzano\textendash Weierstrass Theorem]
Each bounded sequence in $\mathbb{R}^{K}$ has at least one accumulation point or equivalently at least one convergent subsequence.
\end{theorem}
%\textbf{Bolzano\textendash Weierstrass Theorem:} 
%The theorem states that each bounded sequence in $\mathbb{R}^{K}$ has at least one accumulation point or equivalently at least one convergent subsequence.\\
Armed with this theorem, we proceed with our proof by contradiction. Consider that case (b) occurs, we can rewrite the sequence $\boldsymbol{r}^n$ as follows:
$\boldsymbol{r}^n=||\boldsymbol{r}^n||\frac{\boldsymbol{r}^n}{||\boldsymbol{r}^n||}$ where $\frac{\boldsymbol{r}^n}{||\boldsymbol{r}^n||}$ is a bounded sequence. Knowing that $\frac{\boldsymbol{r}^n}{||\boldsymbol{r}^n||}$ in $\mathbb{R}^{K}$ is a bounded sequence, the theorem states that $\frac{\boldsymbol{r}^n}{||\boldsymbol{r}^n||}$ has at least one accumulation point or equivalently at least one convergent subsequence. We denote by $\boldsymbol{\overline{r}}$ one of its accumulations point. Since $F((\boldsymbol{r}^n)_n,\boldsymbol{\rho^*}) \overset{n\rightarrow+\infty}{\longrightarrow} F^*$ then this is still true for any subsequence extracted from $(\boldsymbol{r}^n)_n$. Consider the subsequence corresponding to the accumulation point $\boldsymbol{\overline{r}}$, then for $y\geq0$ which refers to the modulus, $F^*=\displaystyle{\lim_{y \to +\infty}} F(y\boldsymbol{\overline{r}},\boldsymbol{\rho^*})$. We define the set of all states $\Omega=\{(a^j,x^i), j = 1, \ldots, 2^K \;i = 1, \ldots, |I_j|\} $, let $m=\displaystyle{\max_{S\in \Omega}}\: \{\langle x^i,\boldsymbol{\overline{r}}\rangle \}$ and denote $\Gamma=\{S\in \Omega: \langle x^i,\boldsymbol{\overline{r}}\rangle=m\}$ where $\langle\cdot,\cdot\rangle$ refers to the scalar product. Then as $n_k\rightarrow +\infty$, the stationary distribution for a random $S=(a^j,x^i)\in\Omega$ of our chain becomes:
\color{black}
\begin{equation*}
\pi(a^j,x^i;y\boldsymbol{\overline{r}},\boldsymbol{\rho^*})=\frac{\exp(\langle a^j,\boldsymbol{\rho^*}\rangle)\exp(\langle x^i,y\overline{r}\rangle)}{\sum\limits_{j=1}^{2^K}\sum\limits_{i=1}^{|I_j|}\exp(\langle a^j,\boldsymbol{\rho^*}\rangle)\exp(\langle x^i,y\overline{r}\rangle)}
\end{equation*}
By multiplying both the numerator and denominator by the same quantity $\exp(-ym)$, it leads to:
\begin{equation*}
\frac{\exp(y(\langle x^i,\boldsymbol{\overline{r}}\rangle-m +\langle a^j,\boldsymbol{\rho^*}\rangle/y))}{\sum\limits_{j=1}^{2^K}\sum\limits_{i=1}^{|I_j|}\exp(y(\langle x^i,\boldsymbol{\overline{r}}\rangle-m +\langle a^j,\boldsymbol{\rho^*}\rangle/y))}\overset{y\rightarrow+\infty}{\longrightarrow}\frac{\mathbbm{1}\{S\in \Gamma\}}{|\Gamma|}
\end{equation*}
Since $\langle a^j,\boldsymbol{\rho^*}\rangle/y\overset{y\rightarrow+\infty}{\longrightarrow}0$, we are left with the factor $\langle x^i,\overline{r}\rangle-m\leqslant0$ with y tending to infinity. In order to have a non zero numerator we need that factor to be null or in other words $S\in\Gamma$.\\
We can distinguish two cases:\\
- $\Gamma=\Omega$: In this case, all the states share the same maximum value $m$ hence the limiting distribution is simply the uniform distribution over all the state space. In this case, $\pi(a^j,x^i;y\boldsymbol{\overline{r}},\boldsymbol{\rho^*})=\pi(a^j,x^i,0,0)$ and $F^*=F(0,0)$. Therefore, F has a finite minimizer $(0,0)$ which contradicts our assumption\\
- $\Gamma\neq \Omega$: In this case, there exist at least a state $S'=(a^{j'},x^{i'})\notin\Gamma$ such that $\pi(a^{j'},x^{i'};y\boldsymbol{\overline{r}},\boldsymbol{\rho^*})\overset{y\rightarrow+\infty}{\longrightarrow}0$. However, we assumed that $p_{i'j'}>0$ and therefore we can conclude, from the expression of the objective function, that $F^*=+\infty$ which is clearly not minimal since $F(0,0)<+\infty$. This means that case (b) cannot occur\\
The proof that case (c) cannot occur is identical to the preceding one but by taking the sequence $(\boldsymbol{\rho}^n)_n$ into account instead.
The same reasoning can be applied to prove the impossibility of case (d). The two sequences are taken simultaneously and we extract from each a subsequence with their corresponding accumulation points. We will have to define the following quantities: \\
- $m_1=\displaystyle{\max_{S\in \Omega}}\: \{\langle a^j,\boldsymbol{\overline{\rho}}\rangle \}$ and denote $\Gamma_1=\{S\in \Omega: \langle a^j,\boldsymbol{\overline{\rho}}\rangle=m_1\}$ \\
- $m_2=\displaystyle{\max_{S\in \Omega}}\: \{\langle x^i,\boldsymbol{\overline{r}}\rangle \}$ and denote $\Gamma_2=\{S\in \Omega: \langle x^i,\boldsymbol{\overline{r}}\rangle=m_2\}$\\
By multiplying the stationary distribution by $\exp(-m_1y_1-m_2y_2)$, we will end up with:
\begin{equation*}
\pi(a^j,x^i;y_2\boldsymbol{\overline{r}},y_1\boldsymbol{\overline{\rho}})\overset{y_1,y_2\rightarrow+\infty}{\longrightarrow}\frac{\mathbbm{1}\{S\in \Gamma_1\cap\Gamma_2\}}{|\Gamma_1\cap\Gamma_2|}
\end{equation*}

By following the same analysis as before, we will get to the conclusion that case (d) cannot occur and we are left with case (a) which proves the existence of a unique minimizer $(r^*,\rho^*)$ as long as $p_{ij}>0 \:\:\forall\:S\in\Omega$ which brings us back to our original assumptions that it is enough to have $\boldsymbol{\lambda} \in \interior(\Lambda)$ and $\boldsymbol{f} \in \interior(\Theta(\boldsymbol{\lambda}))$
to have a unique finite minimizer $(r^*,\rho^*)$.



%\bibliographystyle{IEEEtran}

\end{document}